\documentclass[onecolumn,amsmath,amssymb,12pt,superscriptaddress,nofootinbib]{revtex4-1}
\pdfoutput=1

\usepackage[english]{babel}
\usepackage{amsthm}
\usepackage{amssymb}
\usepackage{amsmath}
\usepackage{amsthm}
\usepackage{ulem}
\usepackage[]{graphicx}
\usepackage{tensor}
\usepackage{color}
\usepackage{framed}
\usepackage{framed}
\usepackage[bookmarks,linktocpage, colorlinks=true, plainpages = false, citecolor = blue,  linkcolor=blue, urlcolor = blue, filecolor = blue]{hyperref} 
\usepackage{natbib}
\usepackage{dcolumn}
\usepackage{enumerate}
\usepackage{bm}
\usepackage{graphicx}
\usepackage{caption}
\usepackage{subcaption}
\graphicspath{{figures/}}

\theoremstyle{definition}

\newcommand{\sech}{\text{sech}}

\usepackage{tikz}
\usetikzlibrary{decorations.pathmorphing, patterns,shapes}

\input epsf

\begin{document}
\allowdisplaybreaks
\begin{titlepage}
\title{Complex classical paths in quantum reflections and tunneling}
\author{Job Feldbrugge}
\email{job.feldbrugge@ed.ac.uk}
\affiliation{Higgs Centre for Theoretical Physics, James Clerk Maxwell Building, Edinburgh EH9 3FD, UK}
\author{Dylan L. Jow}
\email{d.jow@mail.utoronto.ca}
\affiliation{Canadian Institute for Theoretical Astrophysics, University of Toronto, 60 St. George Street, Toronto, ON M5S 3H8, Canada}
\affiliation{Department of Physics, University of Toronto, 60 St. George Street, Toronto, ON M5S 1A7, Canada}
\affiliation{Dunlap Institute for Astronomy \& Astrophysics, University of Toronto, AB 120-50 St. George Street, Toronto, ON M5S 3H4, Canada}
\author{Ue-Li Pen $^{2,\, 3,\,4,\,}$}
\email{pen@asiaa.sinica.edu.tw}
\affiliation{Institute of Astronomy and Astrophysics, Academia Sinica, Astronomy-Mathematics Building, No. 1, Section 4, Roosevelt Road, Taipei 10617, Taiwan}
\affiliation{Perimeter Institute for Theoretical Physics, 31 Caroline St. North, Waterloo, ON, Canada N2L 2Y5}
\affiliation{Canadian Institute for Advanced Research, CIFAR program in Gravitation and Cosmology}
\begin{abstract}
%\vspace{.25cm}
The real-time propagator of the symmetric Rosen-Morse, also known as the symmetric modified P\"oschl-Teller, barrier is expressed in the Picard-Lefschetz path integral formalism using real and complex classical paths. We explain how the interference pattern in the real-time propagator and energy propagator is organized by caustics and Stoke's phenomena, and list the relevant real and complex classical paths as a function of the initial and final position. We discover the occurrence of singularity crossings, where the analytic continuation of the complex classical path no longer satisfies the boundary value problem and needs to be analytically continued. Moreover, we demonstrate how these singularity crossings play a central role in the real-time description of quantum tunneling.
\end{abstract}

\maketitle 
\end{titlepage}

\tableofcontents

%%%%%%%%%%%%%%%%%%%%%%%%%%
\section{Introduction}
Quantum tunneling is among the most striking phenomena in physics, wherein a classically forbidden transition occurs with a finite probability. The phenomenon was first described by Hund \cite{Hund:1927} and in the following year observed by Landsberg and Mandelstam \cite{Landsberg:1928}. For an overview of the early history of the phenomenon see \cite{Merzbacher:2002}. Quantum tunneling is a central part of modern technology, giving rise to, for example, the tunnel diode, scanning tunneling microscope, and quantum computing. Moreover, it plays an important role in natural phenomena, giving rise to alpha radioactive decay of atomic nuclei, nuclear fusion \cite{Landau:1981}, many biological phenomena including photosynthesis \cite{Renger:2010, Mohseni:2014}, and the ignition of stars \cite{Bohm:1989}. Tunneling may also play a key role in the origin and evolution of life \cite{Trixler:2013} and may ultimately provide a description of the big bang \cite{Lemaitre:1931, Hawking:1982, Vilenkin:1982, Hartle:1983}. Yet surprisingly, quantum tunneling -- well understood in the operator and Euclidean path integral frameworks -- does not yet have a complete description in the real-time Feynman path integral formalism. In this paper, we express the real-time path integral in terms of real and complex classical paths where the complex paths describe classically forbidden behavior. Moreover, we find the complex path responsible for the quantum tunneling and reflection rate.

Quantum tunneling was first discovered using the Schr\"odinger and Heisenberg formulation of quantum mechanics \cite{Landau:1981}. Though very powerful, these formulations of quantum physics obscure its relation to classical mechanics. In the '40s, Wheeler and Feynman proposed the equivalent sum-over-histories formalism. The resulting Feynman path integral is arguably the most elegant formulation of quantum physics, linking quantum physics to classical physics through interference \cite{Feynman:2005, Feynman:1948, Feynman:1965, Feynman:1985}. The probability of a quantum transition follows from the interference over all possible histories, including both on-shell and off-shell real-valued continuous paths interpolating between the initial and final state. Classical evolution arises from constructive interference in the neighborhood of a real classical solution of the equations of motion interpolating between the initial and final state.  Quantum tunneling, when no such real solution exists, is often assumed to correspond to complex classical paths. These complex paths are sometimes known as instantons in the imaginary-time Euclidean path integral \cite{Coleman:1979, Rajaraman:1982}. By considering complex classical paths, the real-time Feynman path integral may bridge the classical and quantum realms, hinting at an explanation of quantum probabilities in terms of classical paths (see for example \cite{Turok:2014, Cherman:2014, Feldbrugge:2017, Matsui:2022}).

However, ever since its inception, the mathematically rigorous definition of the real-time Feynman path integral has been a contentious subject (see for example \cite{Klauder:2003, Klauder:2010}). Consequently, it has been difficult to make the relevance of these complex paths rigorous. In a recent paper, a new definition of the real-time Feynman path integral for quantum mechanics was proposed based on Picard-Lefschetz theory \cite{Feldbrugge:2023}, with close ties to the resurgence program (see for example \cite{Dunne:2015, Dunne:2016, Sueishi:2020}). The path integral is defined over the eigen-thimbles, consisting of sets of complexified off-shell continuous paths, of a set of relevant (complex) classical paths. On each eigen-thimble, the path integral is defined in terms of a functional Wiener integral. In this proposal, classically allowed transitions are dominated by the eigen-thimbles of the real classical paths. Classically forbidden transitions, for which no such real classical path exists, are described by a set of relevant complex classical paths. The relevance of these complex classical paths is formally captured by the steepest descent flow in the space of complex paths. Unfortunately, this condition is currently difficult to implement numerically. This thus raises the question:  when is a complex classical path relevant to the path integral and is such a complex path responsible for quantum tunneling?

In this paper, we study the implementation of this Picard-Lefschetz path integral and the implications of real and complex classical paths in the Rosen-Morse or modified P\"oschl-Teller system. The P\"oschl-Teller potential \cite{Poschl:1933} and its modification to the Rosen-Morse system \cite{Rosen:1932} has been studied for a long time, as it is one of the few exactly solvable quantum systems \cite{Kleinert:1992, Grosche:1998, Grosche:1993, Kleinert:2004}. The potential well is famously reflectionless in specific configurations \cite{Crandall:1983} and its study has even found applications in black hole physics \cite{Ferrari:1984}. For us, the Rosen-Morse system forms an ideal testbed to explore classical paths and its structure, as it enables us to directly verify our results. We find that the Rosen-Morse barrier in quantum mechanics reveals a rich mathematical structure underlying quantum interference that we believe will lead to novel realizations in both real-time relativistic quantum systems and especially in quantum gravity. The analysis of the Rosen-Morse well will follow in an upcoming paper \cite{Feldbrugge:2023Well}.

In section \ref{sec:exact}, we give the closed-form expressions for the real-time and energy propagator of the Rosen-Morse barrier. The classical path analysis in further sections is compared with these exact results. In section \ref{sec:Feynman}, we summarize the Picard-Lefschetz definition of the real-time Feynman path integral and study its implications for the Rosen-Morse barrier. We find a set of real and complex classical paths organized by caustics and Stoke's phenomena that govern the real-time propagator. Moreover, we find a new phenomenon known as the \textit{singularity crossing}, where complex classical paths undergo a singularity after which they fail to solve the boundary value problem. This phenomenon is described in detail in an accompanying letter \cite{Feldbrugge:2023Singular}. We show how the classical paths, their caustics, Stoke's phenomena and singularity crossings have direct manifestations in evolving wave packets. In section \ref{sec:energy}, we analyze the energy propagator, where the caustics are interpreted as turning points. Moreover, we obtain the complex classical paths responsible for quantum tunneling and quantum reflection. As it turns out, the quantum tunneling and quantum reflection can only be reproduced when taking these singularity crossings into account. We summarize our results in section \ref{sec:conclusion}.

%%%%%%%%%%%%%%%%%%%%%%%%%%
\section{The propagator of the Rosen-Morse barrier}\label{sec:exact}
In quantum mechanics, the evolution of a non-relativistic particle with mass $m$ in a Rosen-Morse potential is governed by the Schr\"odinger equation
\begin{align}
    i \hbar \frac{\partial \psi_t}{\partial t}  &= \hat{H} \psi_t\,,\\
    \psi_{t=0}(x) &= \psi_0(x)\,,
\end{align}
with the wavefunction, $\psi_t$, whose amplitude squared, $|\psi_t(x)|^2$, yields the probability density for the particle to be observed at $x$ at time $t$, the initial wavefunction $\psi_0$, the reduced Planck constant, $\hbar$, the Hamiltonian operator,
\begin{align}
    \hat{H} = -\frac{\hbar^2}{2m}\frac{\partial^2}{\partial x^2} + V(x)\,,
\end{align}
and the symmetric Rosen-Morse potential,
\begin{align}
    V(x) = \frac{V_0}{\cosh(x)^2}\,.
\end{align} 
For the barrier, it is natural to consider a positive potential, $V_0 > 0$. In fact, in this paper, we will be a bit more restrictive and consider the Rosen-Morse barrier with $V_0 \geq \frac{\hbar^2}{8m}$ to simplify the analysis. In a companion paper, we consider the Rosen-Morse well with $V_0 <0$ \cite{Feldbrugge:2023Well}. 

%%%%%%%%%%%%%%%%%%%%%%%%%%
\subsection{The spacetime propagator}
The Schr\"odinger equation is solved as the convolution of an initial state, $\psi_0$, with a Green's function, 
\begin{align}
    \psi_T(x_1) =  \int_{-\infty}^\infty G[x_1,x_0;T] \psi_0(x_0)\mathrm{d}x_0\,,
\end{align}
where the Green's function, $G$, captures the dynamics of the Schr\"odinger equation through the differential equation 
\begin{align}
    \left[i\hbar \frac{\partial}{\partial T} - \hat{H}\right] G[x_1,x_0;T] = i \hbar \,\delta^{(1)} (T)\,,
\end{align}
acting on either $x_0$ or $x_1$, with the one-dimensional Dirac delta function $\delta^{(1)}$, and satisfies the boundary condition
\begin{align}
    \lim_{T\to 0} G[x_1,x_0;T] = \delta^{(1)}(x_0-x_1)\,,
\end{align}
ensuring that the initial state is recovered in the limit $T\to 0$ ( \textit{i.e.}, in that limit, we require $\psi_T(x)\to \psi_0(x)$). 

The Rosen-Morse system is one of the few quantum mechanical systems for which the energy eigenstates are known. The energy eigenstates defined by the time-independent Schr\"odinger equation, 
\begin{align}
    \hat{H} \phi_k = \frac{\hbar^2 k^2}{2m} \phi_k\,,
\end{align} 
assume the form
\begin{align}
    \phi^+_k(x) &= \sqrt{\frac{k \sinh(\pi k)}{\cosh(2 \pi k )+ \cosh(2 \pi \nu)}}P_N^{ik}(\tanh x)\,,\\
    \phi^-_k(x) &= \sqrt{\frac{k \sinh(\pi k)}{\cosh(2 \pi k )+ \cosh(2 \pi \nu)}}P_N^{ik}(-\tanh x)\,, 
\end{align}
with positive dimensionless momentum $k$, the associated Legendre function $P_{\sigma}^\mu(x)$, and the constant
\begin{align}
    N=-\frac{1}{2} + \frac{i}{2\hbar} \sqrt{8 m V_0-\hbar^2}\,.
\end{align}
The states $\phi_k^+$/$\phi_k^-$ are the positive/negative momentum states, satisfying the Dirac delta normalization
\begin{align}
    \int \phi^{j_1}_{k_1}(x)\phi^{j_2}_{k_2}(x)^* \mathrm{d}x = \delta_{j_1,j_2}\delta^{(1)}(k_1-k_2)\,.
\end{align}
For a more detailed derivation of the eigenstates see appendix \ref{ap:Teller}. The associated Legendre function, $P_N^\mu$, with degree $N=-\frac{1}{2} + i \nu$ and real $\nu= \frac{1}{2\hbar} \sqrt{8 m V_0-\hbar^2}$ is known as the conical or Mehler function and has many applications in applied mathematics and theoretical physics. Given the eigenstates, we use the spectral representation of the Green's function (see appendix \ref{ap:Green}) to find the Feynman propagator,
\begin{align}
    G[x_1,x_0;T] = \Theta_H(T)\int_{0}^\infty \left[\phi^+_k(x_1)\phi^+_k(x_0)^*+\phi^-_k(x_1)\phi^-_k(x_0)^*\right] e^{-\frac{i \hbar k^2 T}{2m}}\mathrm{d}k\,,
    \label{eq:Green}
\end{align}
with the Heaviside step function $\Theta_H$.  Note that this integral is only conditionally convergent. We define the integral with either a deformation of the integration domain, $\mathbb{R}_{>0}$, in the complex plane using Picard-Lefschetz theory and letting the deformed domain approach the point $e^{-i\pi/4}\infty$ at complex infinity, or with an $i\epsilon$ regulator. For more details on these regulators, see \cite{Feldbrugge:2023}. See appendix \ref{ap:Gevaluation} for a discussion on the numerical evaluation of the integral \eqref{eq:Green}.

%%%%%%%%%%%%%%%%%%%%%%%%%%
\subsection{Reflection and tunneling amplitudes}
The asymptotics of the eigenstates yield the expansion
\begin{align}
    \phi^\pm_k(x) \underset{x \to \pm\infty}{\sim} 
    &\ \mathcal{N}_k \tilde{T}(k) e^{ikx}\,,\\
    \phi^\pm_k(x) \underset{x \to \mp\infty}{\sim}&\
    \ \mathcal{N}_k \left(e^{ikx} + \tilde{R}(k) e^{-ikx}\right)\,,
\end{align}
with the reflection and transmission amplitudes
\begin{align}
    \tilde{R}(k) &= \frac{\Gamma(ik)\Gamma(1-ik+N)\Gamma(-ik-N)}{\Gamma(-ik)\Gamma(-N)\Gamma(N+1)}\,, \\
    \tilde{T}(k) &= \frac{\Gamma(1-ik+N)\Gamma(-ik-N)}{\Gamma(1-ik)\Gamma(-ik)}\,,
\end{align}
expressed in terms of gamma functions $\Gamma$. The normalization constant $\mathcal{N}_k$ is provided in appendix \ref{ap:Teller}. The reflection and tunneling rates assume the form
\begin{align}
    |\tilde{R}(k)|^2&= \cosh(\pi \nu)^2 \sech(\pi (k - \nu)) \sech(\pi (k + \nu))\,,\\
    |\tilde{T}(k)|^2&= \sech(\pi (k - \nu)) \sech(\pi (k + \nu)) \sinh(\pi k)^2\,,
\end{align}
with the imaginary part of $N$ given by $\nu=\text{Im}(N) = \frac{1}{2\hbar} \sqrt{8 m V_0-\hbar^2}$.

\subsection{The energy propagator}
The energy Green's function is the Fourier transform of the spacetime Green's function
\begin{align}
    K[x_1,x_0;E] &= \int_0^\infty G[x_1,x_0;T] e^{i E T/\hbar}\mathrm{d}T\\
    &=i \hbar \int_{0}^\infty \frac{\phi^+_k(x_1)\phi^+_k(x_0)^*+\phi^-_k(x_1)\phi^-_k(x_0)^*}{E - \frac{\hbar^2k^2}{2m}} \mathrm{d}k\,,\label{eq:energydef}
\end{align}
and satisfies the differential equation
\begin{align}
    \left[\hat{H}-E\right] K[x_1,x_0;E] =  -i \hbar \delta^{(1)}(x_0-x_1)\,.
\end{align}
Note that equation \eqref{eq:energydef} is \textit{ab initio} not well defined due to the poles at $k = \pm \sqrt{2mE}/\hbar$. We deform the integral above the right pole for positive $x_1-x_0$ and below for negative $x_1-x_0$ to obtain the Feynman propagator. Alternatively, we can use an $i\epsilon$ prescription. For the Rosen-Morse barrier problem, the energy propagator can be solved in closed-form
\begin{align}
    K[x_1,x_0;E] &= -\frac{im \Gamma(-i k_E - N)\Gamma(- i k_E + N +1)}{\hbar}  P_{N}^{i k_E}(\tanh x_>)P_{N}^{i k_E}(-\tanh x_<)\,,
    \label{eq:KEexact}
\end{align}
with the associated momentum $k_E = \sqrt{2m E}/\hbar$ and the maximum $x_>$ and minimum $x_<$ of the boundary conditions $(x_0,x_1)$ using a Sommerfeld-Watson transformation \cite{Kleinert:1992}.

In the asymptotic region $x_0 \to -\infty$ and $x_1 \to \infty$ with $x_0+x_1=0$, we relate the tunneling amplitude to the energy propagator
\begin{align}
    K[x_1,x_0;E] &\to 
    i \hbar \int_{0}^\infty \frac{ \tilde{T}(k) \left(e^{i k (x_1-x_0)} + \tilde{R}(k) e^{i k (x_0+x_1)}\right) + (x_0 \leftrightarrow x_1)}{E - \frac{\hbar^2k^2}{2m}} \frac{\mathrm{d}k}{2\pi}\\
    &= i \hbar \int_{-\infty}^\infty \frac{ \tilde{T}(k) e^{i k (x_1-x_0)}}{E - \frac{\hbar^2k^2}{2m}} \frac{ \mathrm{d}k}{2 \pi}
    + i \hbar \int_{-\infty}^\infty \frac{ \tilde{T}(k)\tilde{R}(k) }{E - \frac{\hbar^2k^2}{2m}} \frac{ \mathrm{d}k}{2 \pi} \label{eq:ref}\\
    &\sim \tilde{T}\left(k_E\right) K_0[x_1,x_0,E] + \sqrt{\frac{m}{2E}} \tilde{T}\left(k_E\right) \tilde{R}\left(k_E\right) 
\end{align}
where the energy propagator of the free particle is given by
\begin{align}
    K_0[x_1,x_0;E] &= \sqrt{\frac{m}{2E}}e^{i \sqrt{2mE}|x_1-x_0|/\hbar}\\
    &= \sqrt{\frac{m}{2E}}e^{i k_E|x_1-x_0|}\,.
\end{align}
In the limit $x_0\to -\infty$ and $x_1 \to \infty$, we can extract the tunneling amplitude from the exact energy propagator:
\begin{align}
    K[x_1,x_0;E] &= -\frac{im \Gamma(-i k_E - N)\Gamma(- i k_E + N +1)}{\hbar}  P_{N}^{i k_E}(\tanh x_>)P_{N}^{i k_E}(-\tanh x_<)\\
    &\sim \frac{ \Gamma(1- i k_E + N )\Gamma(-i k_E - N)}{\Gamma(1-ik_E)\Gamma(-ik_E)}  \frac{m}{k_E\hbar}e^{ik_E(x_1 -x_0) }\\
    &\sim \tilde{T}(k_E)  K_0[x_1,x_0;E]
\end{align}
using the fact that $P_N^{ik}(\tanh x) \overset{x \to \infty}{\sim} e^{ikx}/\Gamma(1-ik)$ and $x\Gamma(x)=\Gamma(1+x)$. For the Rosen-Morse barrier, the second term in equation \eqref{eq:ref} vanishes. The tunneling rate is thus given by the asymptotic 
\begin{align}
    |\tilde{T}(k_E)|^2 \sim \frac{2E}{m}|K[x_1,x_0;E]|^2
\end{align} 
in the limit $x_0 \to - \infty$ and $x_1 \to \infty$ (or the other way around). In the limit $x_0, x_1 \to \pm \infty$ the exact propagator yields the reflection amplitude
\begin{align}
    K[x_1,x_0;E] \sim K_0[x_1,x_0;E] + \sqrt{\frac{m}{2E}}\tilde{R}(k_E) e^{-i k_E(x_0+x_1)}\,.
\end{align}
The reflection rate thus satisfies the equation
\begin{align}
    |\tilde{R}(k_E)|^2 \sim \frac{2E}{m}\left|K[x_1,x_0; E] - K_0[x_1,x_0;E]\right|^2\,,
\end{align}
when sending both the initial and final position to the left or the right of the potential.

%%%%%%%%%%%%%%%%%%%%%%%%%%
\section{The Feynman path integral}\label{sec:Feynman}
The energy eigenstates yield a very powerful algebraic method to construct the Green's function. However, this technique obscures its relation to classical mechanics and the underlying geometry. Indeed, the energy eigenstates are only known for a handful of quantum mechanical models and are generally not the most practical way to study phenomena in relativistic quantum systems, especially quantum gravity.

Richard Feynman and John Wheeler constructed a powerful method to derive the Green's function based on the physical principle of interference. Instead of summing over solutions of the time-independent Schr\"odinger equation, they propose to \textit{sum over histories}:
\begin{align}
    G[x_1,x_0;T]\ ``\hspace{-1mm}=\hspace{-1mm}" \int_{x(0)=x_0}^{x(T)=x_1} e^{i\int_{t_0}^{t_1}\left[ \frac{1}{2} m \dot{x}^2 - V(x)\right]\mathrm{d}t/\hbar}\ \mathcal{D}x(t)\,,\label{eq:PathIntegral1}
\end{align}
with the time interval $T=t_1-t_0$ \cite{Feynman:1965,Feynman:1985}. In this context, the Green's function is also known as the Feynman propagator. Even though this `integral' has led to enormous progress in our understanding of quantum physics in a vast range of problems, the Feynman path integral in the traditional form has proven mathematically difficult to properly define as a functional integral. In particular, the naive definition faces a number of formal problems:
\begin{itemize}
    \item The infinite product of Lebesgue measures $\mathcal{D}x = \prod_{i} \mathrm{d}x_i$ is not a $\sigma$-measure.
    \item When interpreting \eqref{eq:PathIntegral1} as a proper integral, it can only be conditionally convergent, as the integrand is a pure phase. Conditionally convergent integrals can strongly depend on the employed regularisation procedure as the dominated convergence and Fubini's theorems do not apply.
    \item When interpreting \eqref{eq:PathIntegral1} as a proper integral, the paths over which it has support consist of almost nowhere differential paths, in accordance with the Heisenberg uncertainty principle, making the kinetic term in the exponent ambiguous.
\end{itemize}
For more details on the mathematical subtleties of the real-time path integral see \cite{Klauder:2010} and references therein. In a recent paper, a new definition of the Feynman path integral based on Picard-Lefschetz theory was constructed, 
\begin{align}
    G[x_1,x_0;T] =  \Theta_H(T)\sqrt{\frac{m}{2 \pi i \hbar T}} \sum_{x_C} e^{iS_C/\hbar} \int_{\mathcal{J}_C} e^{i \theta_C} \mathrm{d} \mu_C\,,\label{eq:PathIntegral2}
\end{align}
where classical theory plays a central organizing role \cite{Feldbrugge:2023}. We extend the integration domain, consisting of continuous real-valued paths interpolating the initial $x_0$ and final position $x_1$, to complex-valued paths. In this space, we deform the original integration domain to a set of thimbles associated with classical paths solving the corresponding boundary value problem. 

In particular, the definition consists of a sum over a set of relevant (complex) classical paths, $x_C$, for which we construct the descent thimble, $\mathcal{J}_C$, and define a well-defined $\sigma$-measure, $\mu_C$, using Wiener integration theory. A classical path, $x_C$, is relevant when it can be deformed to the original integration domain of real-valued continuous paths following an ascent flow. Real solutions to the boundary value problem are automatically relevant. Complex solutions may or may not play a role in the path integral. For each classical path, we include the associated classical action, $S_C= \int(m \dot{x}_C^2/2 -V(x_C))\mathrm{d}t$, and the phase factor, $e^{i\theta_C}$, on the space of complex continuous paths capturing the geometry of the thimble, $\mathcal{J}_C$. In the semi-classical limit, $\hbar \to 0$, the measure becomes tightly focused on the classical path, $x_C$, and the term $e^{i\theta_C}$ tends to the well-known semi-classical ``Maslow'' phase factor. For more details on this construction see \cite{Feldbrugge:2023}.

The occurrence of relevant complex classical paths in the path integral has been discussed for a long time, while it has to our knowledge not been explicitly implemented in this way using Picard-Lefschetz theory. Indeed, already in 1981, L. S. Schulman in his insightful book on the path integral method \cite{Schulman:2012} foreshadows our work stating: \textit{``It may happen that \textit{no} solution exists but that by analytic continuation complex times or complex paths may be found that are stationary. Such `complex paths' suggest a relation to geometric diffraction theory, and were it not for the formal nature of our stationary phase approximation for the path integral the way would be open to a good deal of useful mathematics.''}

%%%%%%%%%%%%%%%%%%%%%%%%%%
\subsection{Classical paths}
The classical paths for a particle of mass $m$ in a symmetric xRosen-Morse potential is governed by the action
\begin{align}
    S[x] = \int \left(\frac{1}{2} m \dot{x}^2 - \frac{V_0}{\cosh(x)^2}\right)\mathrm{d}t\,,
\end{align}
leading, via the Euler-Lagrange equations, to the equation of motion in the form of Newton's second law 
\begin{align}
    m \ddot{x}(t) = -V'(x(t)) = \frac{2 V_0\tanh(x(t))}{\cosh^2(x(t))}\,.
\end{align}
As the Lagrangian does not explicitly depend on time, the equation of motion conserves the energy
\begin{align}
    E &=\frac{m\dot{x}(t)^2}{2}+\frac{V_0}{\cosh^2(x(t))}\\
    &= \frac{m v_0^2}{2} + \frac{V_0}{\cosh^2(x_0)}\,,
\end{align}
with the initial position $x(0)=x_0$ and velocity $\dot{x}(0)=v_0$. The classical paths thus satisfy the integral equation
\begin{align}
    x_C(t) &= \pm \int \sqrt{\frac{2}{m}\left[E - \frac{V_0}{\cosh^2(x_C(t))}\right]}\ \mathrm{d}t\,,
\end{align}
yielding the explicit solution
\begin{align}
    \sinh(x_C(t)) = c_1 \sqrt{\frac{E-V_0}{E}} \sinh\left[ \sqrt{\frac{2E}{m}}(t-C) \right]\,.\label{eq:sol1}
\end{align}
For the initial value problem $x(0)=x_0$ and $\dot{x}(0)=v_0$, we explicitly solve for the shift parameter
\begin{align}
    C = c_2 \sqrt{\frac{m}{2E}} \sinh^{-1}\left[ \sqrt{\frac{E}{E-V_0}} \sinh(x_0) \right]\,.
\end{align}
The expressions for the classical paths are well-defined up to two signs $c_1,c_2 = \pm1$ associated with the choice of Riemann sheets of the two square roots and the arcsinh function. Given the choice of the Riemann sheets and the signs $c_1,c_2$, equation \eqref{eq:sol1} is a closed-form solution of the classical paths of the Rosen-Morse theory. The classical action assumes the corresponding closed form
\begin{align}
    S[x_C(t)] =  ET-\sqrt{2m V_0}\tanh^{-1}\left[\sqrt{\frac{V_0}{E}} \tanh\left( \sqrt{\frac{2 E}{m}}(t-C) \right)\right]\bigg|_{t=0}^T\,.
    \label{eq:classicalAction}
\end{align}

Note that both the arcsinh and arctanh functions are multi-valued. On the other hand, the initial value problem is well-defined and has a unique solution. As we will see, the Riemann sheets and branch cuts of the arcsinh and arctanh functions will play a pivotal role in the propagator and quantum tunneling.

%%%%%%%%%%%%%%%%%%%%%%%%%%
\subsection{Real classical paths and caustics}
\begin{figure}
    \centering
    \begin{subfigure}[b]{0.47\textwidth}
        \includegraphics[width =\textwidth]{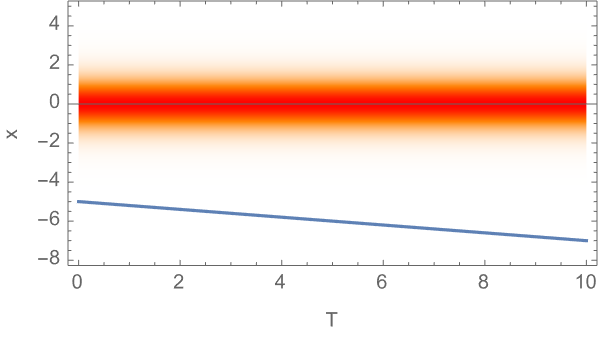}
        \caption{$x_1=-7$}
    \end{subfigure}
    \begin{subfigure}[b]{0.47\textwidth}
        \includegraphics[width =\textwidth]{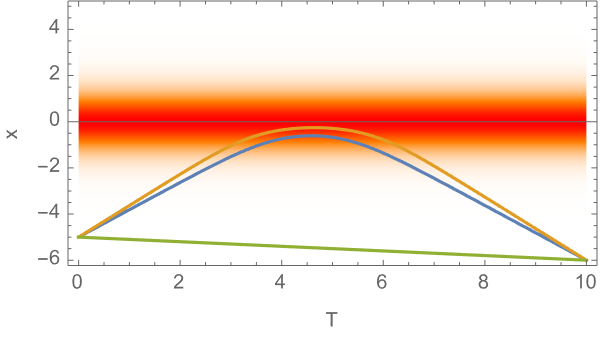}
        \caption{$x_1=-6$}
    \end{subfigure}\\
    \begin{subfigure}[b]{0.47\textwidth}
        \includegraphics[width =\textwidth]{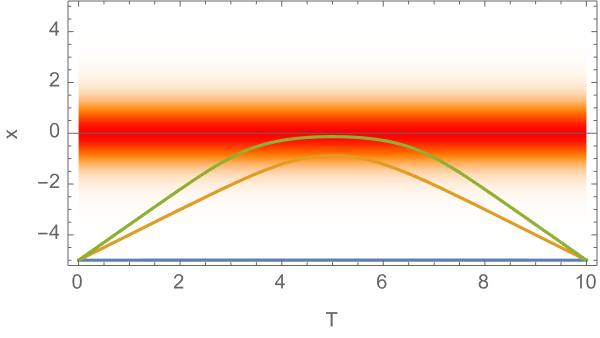}
        \caption{$x_1=-5$}
    \end{subfigure}
    \begin{subfigure}[b]{0.47\textwidth}
        \includegraphics[width =\textwidth]{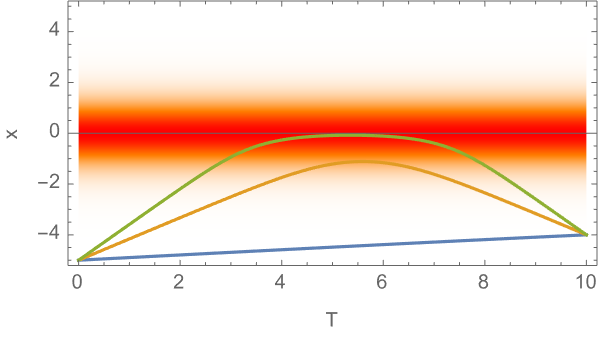}
        \caption{$x_1=-4$}
    \end{subfigure}\\
    \begin{subfigure}[b]{0.47\textwidth}
        \includegraphics[width =\textwidth]{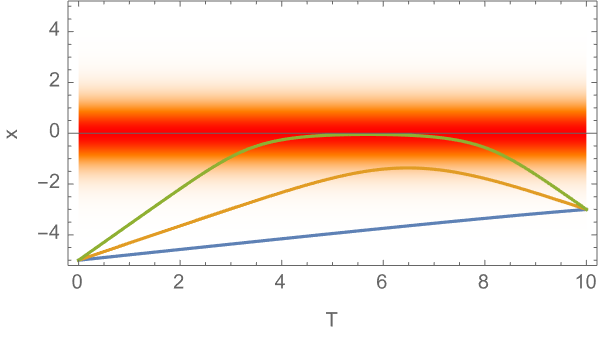}
        \caption{$x_1=-3$}
    \end{subfigure}
    \begin{subfigure}[b]{0.47\textwidth}
        \includegraphics[width =\textwidth]{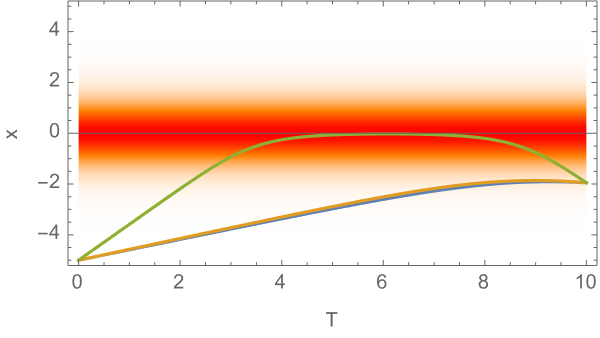}
        \caption{$x_1=-1.95$}
    \end{subfigure}\\
    \begin{subfigure}[b]{0.47\textwidth}
        \includegraphics[width =\textwidth]{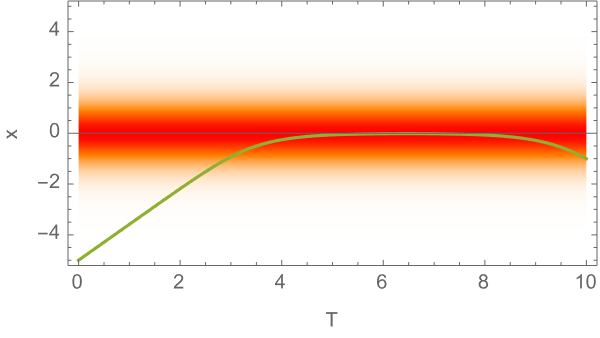}
        \caption{$x_1=-1$}
    \end{subfigure}
    \begin{subfigure}[b]{0.47\textwidth}
        \includegraphics[width =\textwidth]{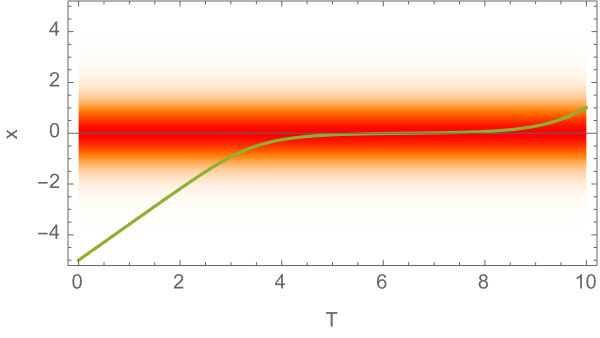}
        \caption{$x_1=1$}
    \end{subfigure}
    \caption{Real paths propagating between $x_0=-5$ and several $x_1$ in time $T=10$ of a particle with mass $m=1$ interacting with a barrier of strength $V_0=1$.}\label{fig:paths}
\end{figure}

\begin{figure}
    \centering
    \includegraphics[width = 0.75\textwidth]{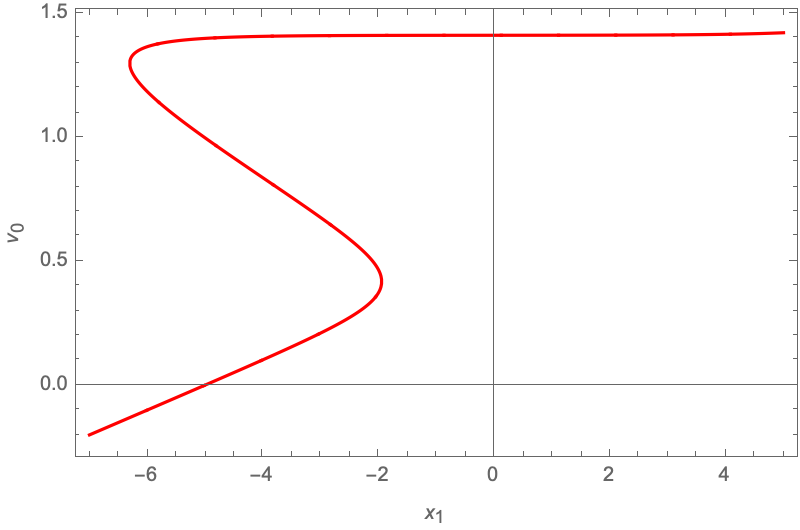}
    \caption{Real initial velocities $v_0$ as a function of the final position $x_1=x(x_0,v_0,T)$ for a particle with mass $m=1$ starting at the initial position $x_0=-5$, evolving for a time $T=10$, and interacting with a barrier of strength $V_0=1$.}\label{fig:RealVelocities}
\end{figure}

The initial value problem always yields a unique solution. This is not true for the boundary value problem, \textit{i.e.}, a final position, $x_1$, may be reached with multiple initial velocities, $v_0$. For example, consider a particle starting at $x_0=-5$ while we vary the final position at a fixed propagation time, $T=10$ (see fig.\ \ref{fig:paths}). For the final position $x_1=-7$, there exists only a single real classical path moving to the left. For larger $x_1$, there exist three real solutions. We find a direct path and two bouncing paths. The bouncing path with the smallest initial velocity approaches the barrier and quickly turns around. The other bouncing path spends a lot of time near the top of the barrier before approaching the final position. At $x_1=-1.95$, two of the three solutions merge and cease to exist. For larger but negative $x_1$, only the path spending a lot of time near the top of the barrier exists. It is not possible to reach $x_1$ directly in the set time. For positive $x_1$, there exists only a single real solution to the boundary value problem, spending some time near the top of the barrier before approaching $x_1$. The merger of real classical paths is most easily observed in terms of the initial velocities as a function of the final position (see fig.\ \ref{fig:RealVelocities}). At $x_1=-6$, two bouncing solutions emerge. At $x_1=-1.95$, the direct solution merges with the bouncing solution with the smaller initial velocity. The bouncing solution with the larger initial velocity persists and turns into the classical path moving across the barrier.

\begin{figure}
    \centering
    \begin{subfigure}[b]{0.49\textwidth}
    \includegraphics[width=\textwidth]{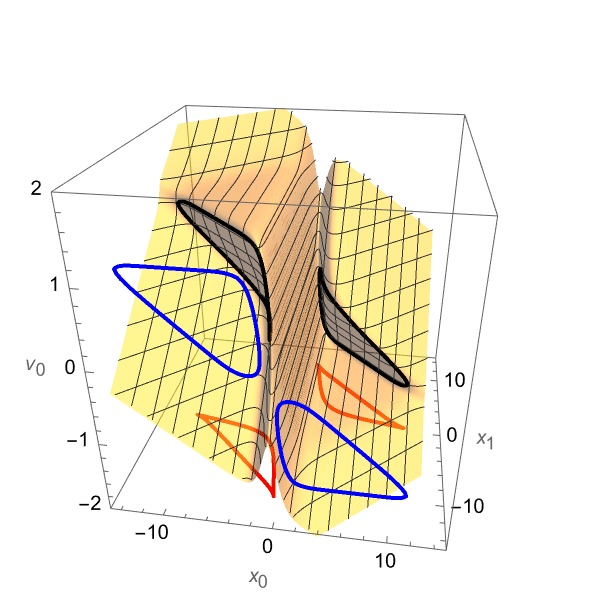}
    \end{subfigure}~
    \begin{subfigure}[b]{0.49\textwidth}
    \includegraphics[width=\textwidth]{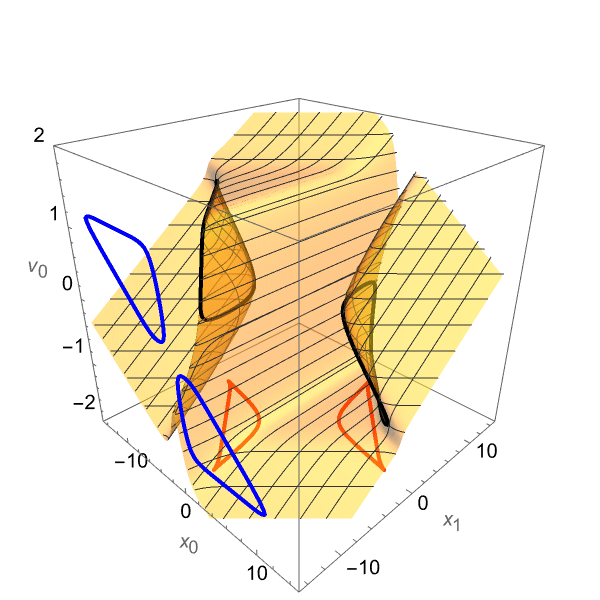}
    \end{subfigure}
    \caption{The classical path surface $\mathcal{M}$ in the space of the initial positions and velocities, and the final positions, with the critical curve (blue), the caustic curve (red), and the curve where the surface $\mathcal{M}$ folds over (black) viewed from two perspectives.}
    \label{fig:phasespace}
% \end{figure}
% \begin{figure}
%     \centering
    \begin{subfigure}[b]{0.49\textwidth}
        \includegraphics[width =\textwidth]{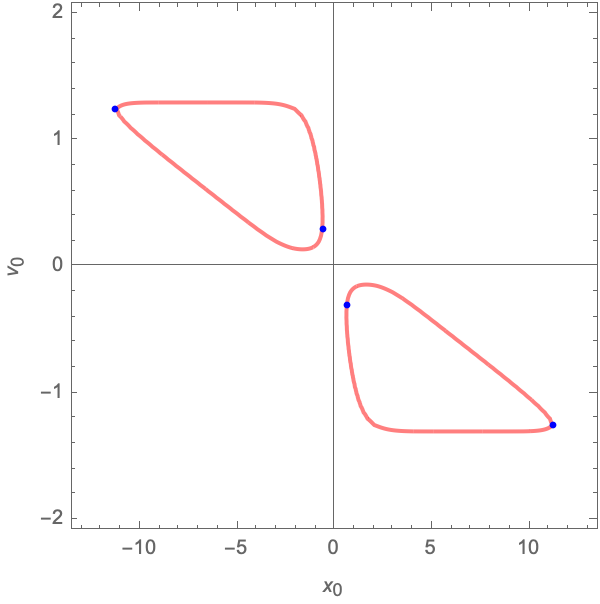}
    \end{subfigure}~
    \begin{subfigure}[b]{0.49\textwidth}
        \begin{tikzpicture}
            \node[anchor=south west,inner sep=0] at (0,0) {\includegraphics[width = \textwidth]{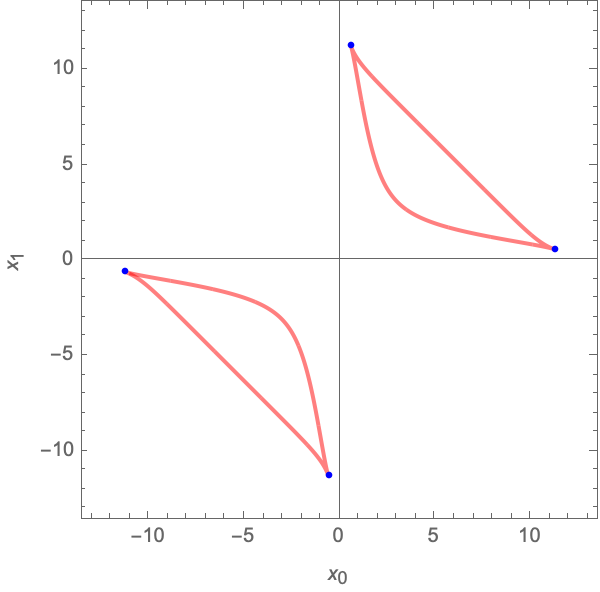}};
            \node[] at (4.55,4.55) {\Large $1$};
            \node[] at (3.4,3.4) {\Large $3$};
            \node[] at (5.6,5.6) {\Large $3$};
        \end{tikzpicture}
    \end{subfigure}
    \caption{The fold caustic (red) and cusp caustic (blue) on the critical (left) and caustic curves (right) of the Teller potential for a particle of mass $m=1$, evolving for a propagation time $T=10$ interacting with a barrier of strength $V_0=1$.}\label{fig:Caustics}
\end{figure}

This behavior can most easily be understood geometrically. Given the initial position, $x_0$, and velocity, $v_0$, we evaluate the associated final position, $x_1$, reached after a time $T$. The set of points $\mathcal{M}=\{(x_0,v_0,x_1)\,|\,x_C(x_0,v_0,T)=x_1\}$ forms a two-dimensional surface representing the space of real classical paths (see fig.\ \ref{fig:phasespace}). The solutions to the boundary value problem with $x(0)=x_0$ and $x(T)=x_1$ emerge as the intersections of the surface $\mathcal{M}$ with the line $\mathcal{L} = \{(x_0,\lambda,x_1)|\lambda \in \mathbb{R}\}$. When the potential is free of singularities (on the real line), the surface $\mathcal{M}$ consists of one piece and is smooth. The boundary value problem will always have an odd number of solutions \cite{Arnold:1978, Arnold:1984}. In particular, for the Rosen-Morse tunneling problem, either one or three real classical paths satisfy the boundary conditions. 

\begin{figure}
    \centering
    \begin{tikzpicture}
        \node[anchor=south west,inner sep=0] at (0,0) {\includegraphics[width = 0.5 \textwidth]{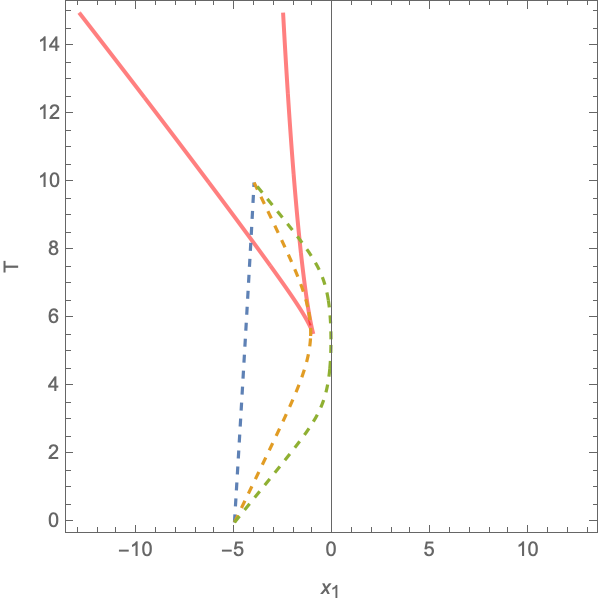}};
        \node[] at (5.5,3.7) {\Large $1$};
        \node[] at (3.4,6.5) {\Large $3$};
    \end{tikzpicture}
    \caption{The evolution of the caustics in time (red) for a particle of mass $m=1$ starting at the initial position $x_0=-5$. In addition, we plot the three real classical paths (the dashed curves) moving from the initial position $x_0=-5$ to the final position $x_1=-4$ at time $T=10$.}
    \label{fig:Caustics_time}
\end{figure}

The $(x_0,x_1)$-space can be partitioned into regions with a distinct number of real classical paths. These regions are bounded by caustics, where the tangent space of the surface $\mathcal{M}$ becomes colinear with the vertical line, $\mathcal{L}$. Explicitly, we can define the critical curve as the points for which the surface $\mathcal{M}$ folds over. In the $(x_0,v_0)$-plane, we define the critical curve as
\begin{align}
    \mathcal{C}=\left\{(x_0,v_0)\,|\, \partial x_C(x_0,v_0;T)  / \partial v_0 = 0\right\}.
\end{align}
The caustic curve is the pushforward of the critical curve to the $(x_0,x_1)$-plane,
\begin{align}
    x_C(\mathcal{C}) = \left\{ (x_0,x_1)\,|\, x_1 = x_C(x_0,v_0;T)\text{ with } (x_0,v_0) \in \mathcal{C}\right\}\,.
\end{align}
See fig.\ \ref{fig:Caustics} for the relevant critical and caustic curves of the Rosen-Morse barrier. Inside the caustic curves, the boundary value problem has three real solutions. Outside the caustic curves, there exists only a single real solution. As we approach the caustic curve from the inside, multiple classical paths merge. As we cross the caustic curve, two of the merging paths will cease to be real. Note that the derivative of the final position with respect to the initial momentum is proportional to the Jacobian $J=\frac{\partial x_1}{\partial p_0}= \left(-\frac{\partial S_C}{\partial x_0 \partial x_1}\right)^{-1}$ with the initial momentum, $p_0$, and the classical action, $S_C$, related by the Hamilton-Jacobi equation, $p_0 = -\frac{\partial S_C}{\partial x_0}$ (see for example the textbook \cite{Schulman:2012}). The Jacobian will play an important role when working with the saddle point approximation of the Feynman path integral. In time, the region with three real classical paths expands (see fig.\ \ref{fig:Caustics_time}). Note that each classical path moving from $x_0=-5$ into the caustic region passes once through a caustic (or focal point). Classical paths moving to points outside the caustic region will not intersect the caustic curve at some point in time. The three real classical paths merge when moving towards the cusp points. These points correspond to kinks in the caustic curve (formally known as the fold). For more details on higher-order caustics and their definitions in terms of dynamical systems we refer to \cite{Feldbrugge:2018JCAP}.

\begin{figure}
    \centering
    \begin{subfigure}[b]{0.49\textwidth}
        \includegraphics[width =\textwidth]{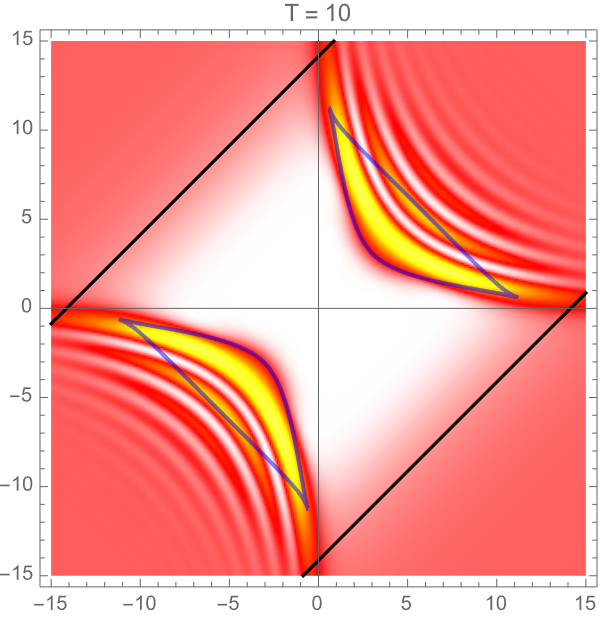}
        \caption{$|G[x_1,x_0;T]|^2$}
    \end{subfigure}
    \begin{subfigure}[b]{0.49\textwidth}
        \includegraphics[width =\textwidth]{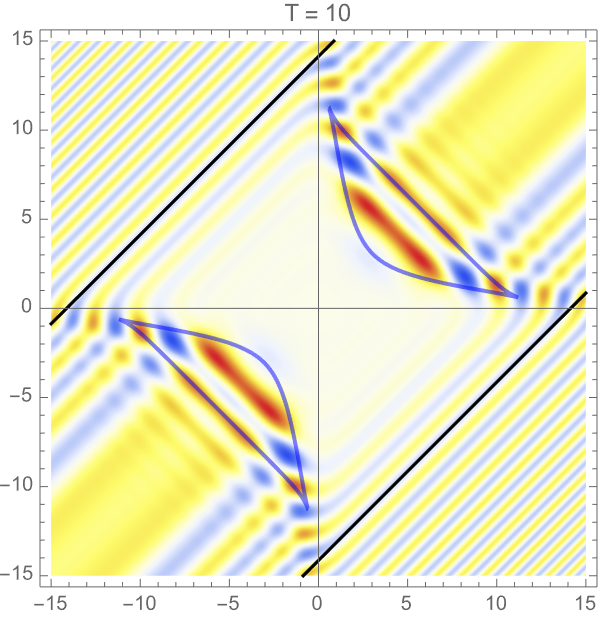}
        \caption{$Re[G[x_1,x_0;T]]$}
    \end{subfigure}
    \caption{The Feynman propagator as a function of the initial and final position $(x_0,x_1)$ for mass $m=1$, time $T=10$, barrier strength $V_0=1$ and reduced Planck constant $\hbar = 1/2$.}
    \label{fig:exactProp}
\end{figure}

Caustics are not only integral to classical mechanics but also organize the propagator in the quantum theory (see fig.\ \ref{fig:exactProp}). The propagator, obtained with the equation \eqref{eq:Green}, consists of an interference pattern in the caustic curve radiating out to the single image region. When $\frac{m (x_0-x_1)^2}{2 T^2} \gg V_0$, the evolution is dominated by the kinetic term, and the propagator approaches the free theory, \textit{i.e.},
\begin{align}
    G_0[x_1,x_0;T] \sim  \Theta_H(T)\sqrt{\frac{m}{2\pi i \hbar T}} e^{i \frac{m (x_1-x_0)^2}{2T\hbar}}\,.
\end{align}
The Feynman path integral shows remarkable similarities with the amplitude of lensed images in wave optics, as described by the Kirchhoff-Fresnel integral \cite{Berry:1980, Feldbrugge:2023AnPhy}. At the fold caustic, two classical paths merge. At the cusp caustics, three classical paths coalesce (for more details on the nature of the different caustics we refer to \cite{Feldbrugge:2018JCAP} and \cite{Feldbrugge:2023AnPhy}). The presence of caustics in the path integral was realized many years ago while catastrophe theory was being developed \cite{Coyne:1972, Schulman:1975, Dewitt:1976, Dewitt:1976b,  Levit:1978, Dewitt:2005}, but has for some reason not become part of the standard treatment of path integral methods (with the notable exception of \cite{Schulman:2012}). In particular, while formula \eqref{eq:Green} for the real-time propagator of the Rosen-Morse barrier system has been known for a long time, fig.\ \ref{fig:exactProp} is to our knowledge the first visualization of the propagator. Caustics organize the qualitative behavior of the real-time propagator for any non-linear quantum system. It is at present unclear how caustics are realized in Euclidean path integral methods.

\begin{figure}
    \centering
    \begin{subfigure}[b]{0.49\textwidth}
        \includegraphics[width =\textwidth]{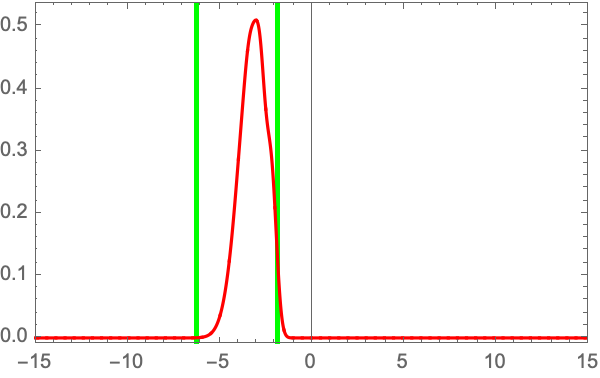}
        \caption{$\sigma=0.5$}
    \end{subfigure}
    \begin{subfigure}[b]{0.49\textwidth}
        \includegraphics[width =\textwidth]{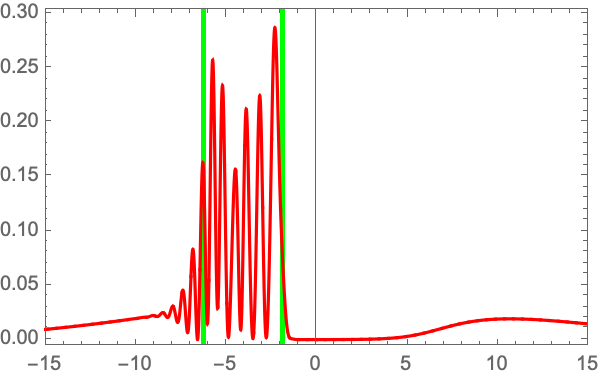}
        \caption{$\sigma=0.05$}
    \end{subfigure}\\
    \begin{subfigure}[b]{0.49\textwidth}
        \includegraphics[width =\textwidth]{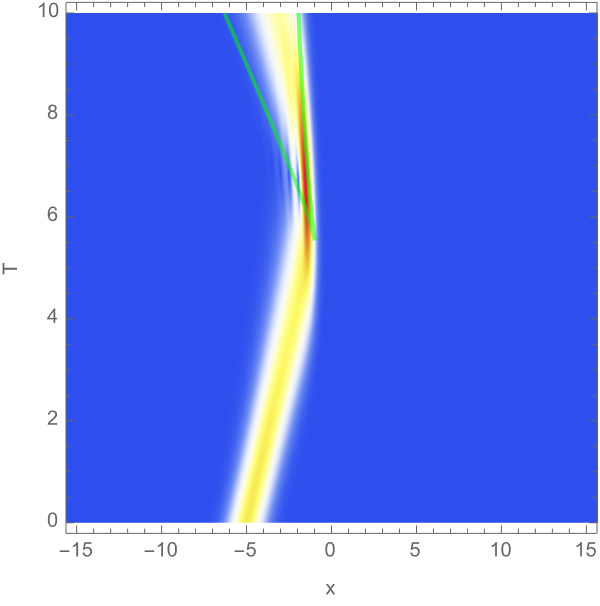}
        \caption{$\sigma=0.5$}
    \end{subfigure}
    \begin{subfigure}[b]{0.49\textwidth}
        \includegraphics[width =\textwidth]{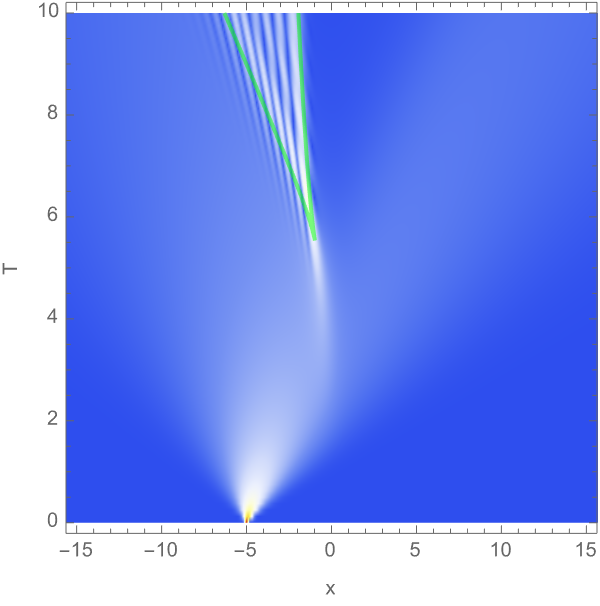}
        \caption{$\sigma=0.05$}
    \end{subfigure}
    \caption{Caustics in the evolution of a Gaussian wave packet for a particle with mass $m=1$, evolving for a time $T=10$, interacting with a barrier of strength $V_0=1$, and with the reduced Planck constant $\hbar=0.1$. The upper panels show the probability densities at $T=10$ with the caustics represented by the green lines. The lower panels show the evolution of the probability density with the caustics in green. The left panels show the evolution of a Gaussian wave packet with a broad peak in position and a narrow peak in momentum. The right panels show the evolution of a Gaussian wave packet that is narrowly peaked in position and broadly peaked in momentum.}
    \label{fig:Schrodinger}
\end{figure}

Caustics are not only central to the Feynman propagator but can also be observed in solutions of the Schr\"odinger equation. Consider for example the evolution of the Gaussian initial state, 
\begin{align}
    \psi_0(x_0) = \frac{1}{\sqrt[4]{2 \pi \sigma_0^2}} e^{-\frac{(x_0-\bar{x}_0)^2}{4 \sigma_0^2} + i \frac{\bar{p}_0 x_0}{\hbar}}\,,
\end{align}
with the mean initial position $\bar{x}_0$, momentum $\bar{p}_0$ and spread $\sigma_0$. The Gaussian state describes a particle with a Gaussian position and momentum distribution
\begin{align}
    p_x(x_0) &=\frac{1}{\sqrt{2 \pi \sigma_0^2}} e^{-\frac{(x_0-\bar{x}_0)^2}{2 \sigma_0^2}}\,,\\
    p_k(p_0) &= \sqrt{\frac{2 \sigma_0^2}{\pi \hbar^2}}e^{-\frac{2 \sigma_0^2 (p_0 - \bar{p}_0)^2}{\hbar^2}}\,,
\end{align}
saturating the Heisenberg uncertainty relation $\Delta x \Delta p = \frac{\hbar}{2}$. A Gaussian state with the mean position $\bar{x}_0=-5$ and momentum $\bar{p}_0=0.7$, and spread $\sigma_0=0.5$ evolves in time $T=10$ into a distribution for which the reflected part resembles a simple bell-shape (see fig.\ \ref{fig:Schrodinger}). The particle, starting roughly at $x=-5$ with a sharply peaked distribution in momentum space, bounces off the barrier and evolves to a state around $x=-3.5$. This is in agreement with the classical path rolling only slightly up the hill (see the central classical paths in fig.\ \ref{fig:RealVelocities}). When we decrease the spread in position to $\sigma_0=0.05$ and increase the uncertainty in momentum space accordingly, we become sensitive to all three real classical paths. The initial state evolves into a highly oscillatory function bounded by the two caustics corresponding with the initial position $x_0=-5$ (see fig.\ \ref{fig:Schrodinger}). The classical caustics directly manifest themselves in the evolution of the wavefunction. We can also find the manifestation of the cusp caustic in the evolution of the wavefunction, in addition to the two fold caustics (see the bottom panels of fig.\ \ref{fig:Schrodinger}).  At time $T\approx5.5$ we observe a cusp caustic, at which the probability is amplified. As the system evolves the interference region spreads. By decreasing the reduced Planck constant, the initial state becomes sharply peaked in momentum space, allowing one to select the bouncing solution rolling only slightly up the hill. Then, by varying the mean initial momentum, $\bar{p}_0$, and spread, $\sigma$, we can probe the three bouncing classical paths from the Feynman propagator individually away from the caustics where they coalesce.

%%%%%%%%%%%%%%%%%%%%%%%%%%
\subsection{Complex classical paths, Stoke's phenomena, and singularity crossings}
The caustics of the classical theory organize the structure of the Feynman propagator. As real classical paths emerge or annihilate, the character of the interference pattern changes. This is particularly clear in the semi-classical limit, $\hbar \to 0$, where the measure in definition \eqref{eq:PathIntegral2} becomes tightly focused on the relevant classical paths, $x_C$, yielding the saddle-point approximation of the real-time path integral, 
\begin{align}
    G_{WKB}[x_1,x_0;T] &= \Theta_H(T) \sqrt{\frac{i}{2\pi \hbar}}
    \sum_{x_C}\sqrt{\frac{\partial^2 S_C}{\partial x_0 \partial x_1}}  e^{i S_C/\hbar}\\
    &= \Theta_H(T)\sqrt{\frac{i}{2\pi \hbar}}\sum_{x_C}\sqrt{-\frac{\partial p_0}{\partial x_1}}  e^{i S_C/\hbar}\\
    &= \Theta_H(T)\sqrt{\frac{1}{2\pi \hbar i}}\sum_{x_C}\left(\frac{\partial x_1}{\partial p_0}\right)^{-1/2}  e^{i S_C/\hbar}\,,
    \label{eq:WKB}
\end{align}
using the Hamilton-Jacobi equation $p_0 = -\frac{\partial S_C}{\partial x_0}$, assuming the paths to be sufficiently distinct.
% \begin{align}
%     (x_{C_1}-x_{C_2})^2 \delta^2S \gg \hbar\,,
% \end{align}
% for all relevant classical paths $x_{C_1}$ and $x_{C_2}$. 
The square root of the Van Vleck-Morette determinant $\partial^2 S_C/\partial x_0 \partial x_1$ can be ambiguous as it is at first not clear which sign we should select. In particular, as the reciprocal $\left(\partial^2 S_C/\partial x_0 \partial x_1\right)^{-1}$ vanishes at the caustics the square root can move between Riemann sheets. It can be shown that the functional determinant arises as the product of eigenvalues of the second-order variation of the action $\delta^2 S$ associated with the classical path $x_C$. In this language, we find that each negative eigenvalue leads to a phase $e^{i \pi /2}$. Or equivalently, we can write $\sqrt{\partial^2 S_C/\partial x_0 \partial x_1} = \sqrt{\left| \partial^2 S_C/\partial x_0 \partial x_1\right|} e^{i n \pi /2}$ where $n$ is the number of focal points of the classical path $x_C$ where the van Vleck-Morette determinant vanishes. The integer $n$ is sometimes known as the Maslow index. For complex paths, we determine the correct Riemann sheet by tracking the evolution of $\partial^2 S_C/\partial x_0 \partial x_1$ as we let the time $t$ run from $0$ to $T$. Each time the functional determinant circles the origin, it receives a sign. The saddle-point approximation is completely governed by relevant classical paths. The approximation diverges at caustics, $x_C(\mathcal{C})$, where $\partial x_1 / \partial p_0 = 0$ and two classical paths coalesce. For a derivation and more details of the saddle point approximation see for example the textbook \cite{Schulman:2012} and references therein.

\begin{figure}
    \centering
    \begin{subfigure}[b]{0.49\textwidth}
        \includegraphics[width =\textwidth]{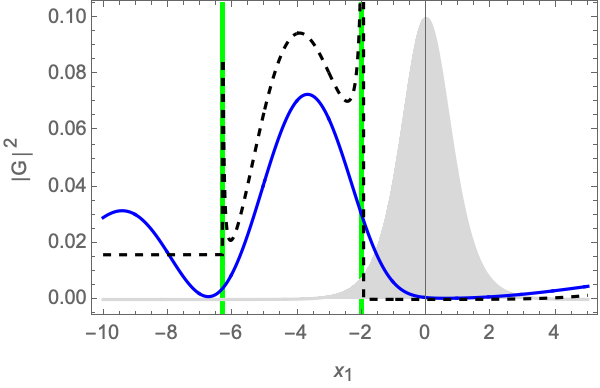}
        \caption{$\hbar=1$}
    \end{subfigure}
    \begin{subfigure}[b]{0.49\textwidth}
        \includegraphics[width =\textwidth]{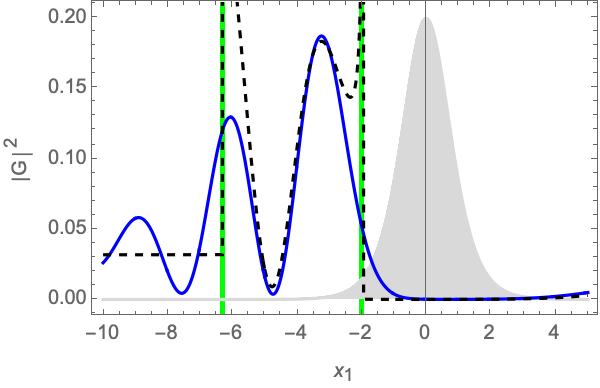}
        \caption{$\hbar=0.5$}
    \end{subfigure}
    \begin{subfigure}[b]{0.49\textwidth}
        \includegraphics[width =\textwidth]{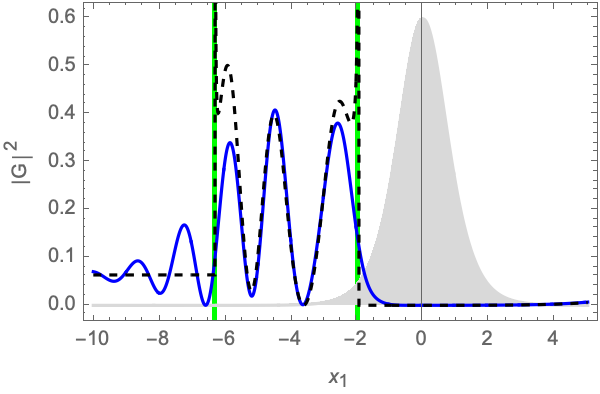}
        \caption{$\hbar=0.25$}
    \end{subfigure}
    \begin{subfigure}[b]{0.49\textwidth}
        \includegraphics[width =\textwidth]{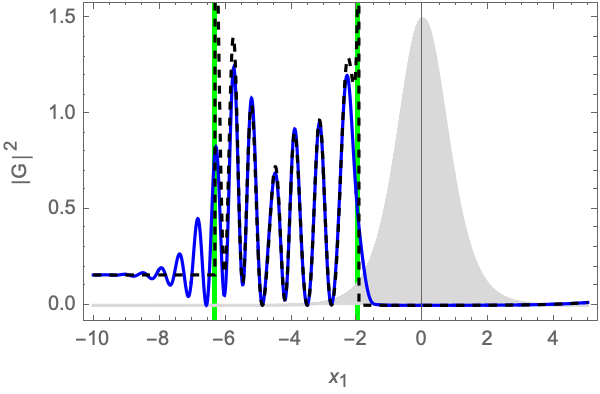}
        \caption{$\hbar=0.1$}
    \end{subfigure}
    \caption{The Feynman propagator as a function of the final position $x_1$ for the initial position $x_0=-5$, mass $m=1$ time $T=10,$ and barrier strength $V_0=1$ for $\hbar=0.1, 0.25,0.5,1$. The exact (blue) and the saddle point approximation based on the real classical paths (black). The caustics are marked by green vertical lines.}
    \label{fig:WKB_line_real}
\end{figure}

The saddle point approximation, of course, reiterates the main question of this paper: which classical paths are relevant to the path integral and which ones should be ignored? Picard-Lefschetz theory proves that a classical path is formally relevant if and only if it can be deformed to the original integration domain by the steepest ascent flow. This theorem indicates that all real classical paths are relevant to the path integral. For this reason, let's compare the propagator for fixed initial position $x_0=-5$ and time $T=10$ with the saddle approximation including only the real classical paths (see fig.\ \ref{fig:WKB_line_real}). The \textit{real saddle-point approximation} excellently reproduces the interference pattern inside the caustic region in the semi-classical limit, $\hbar \to 0$, away from the caustics. The oscillations emerge from the interference of the three real classical paths (see fig.\ \ref{fig:paths}). At the caustic, the propagator smoothly transitions from an oscillatory to a nonoscillatory phase, as in the Airy function, $Ai(x)$ (a classic result in the study of asymptotics and resurgence theory). The real saddle point approximation diverges at the caustic and fails to capture the transition to the region with only a single real classical path.

\begin{figure}
    \centering
    \begin{subfigure}[b]{0.49\textwidth}
        \begin{tikzpicture}
            \node[anchor=south west,inner sep=0] at (0,0) {\includegraphics[width = \textwidth]{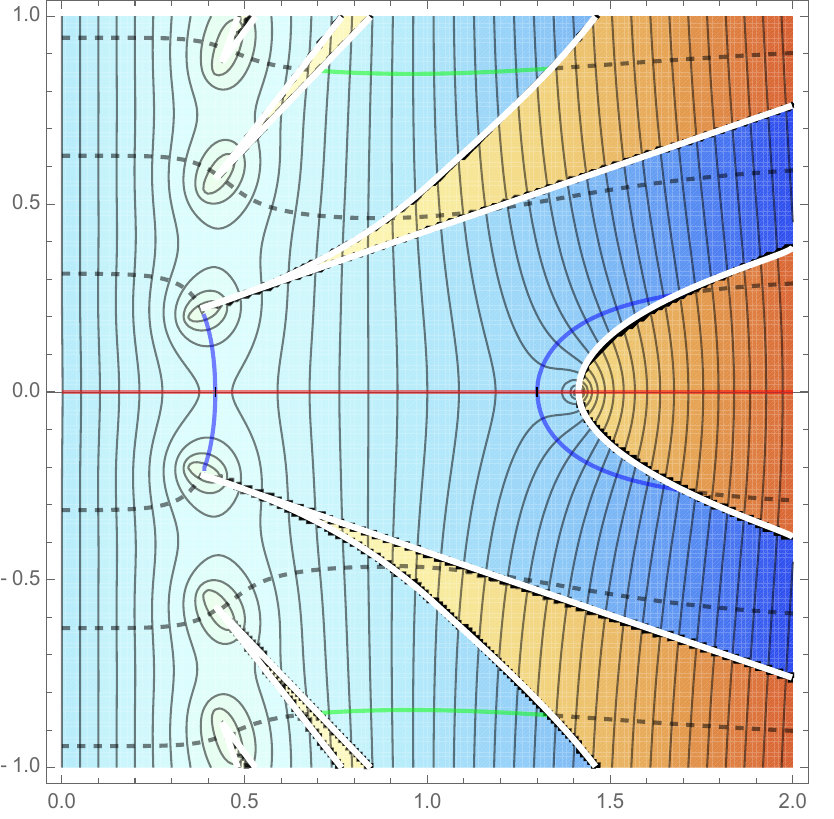}};
            \node[] at (1.3,4.375) {\large I};
            \node[] at (4,4.375) {\large II};
            \node[] at (7,4.375) {\large III};
        \end{tikzpicture}
    \caption{$x_1$}\label{fig:Tracks}
    \end{subfigure}
    \begin{subfigure}[b]{0.49\textwidth}
        \begin{tikzpicture}
            \node[anchor=south west,inner sep=0] at (0,0) {\includegraphics[width = \textwidth]{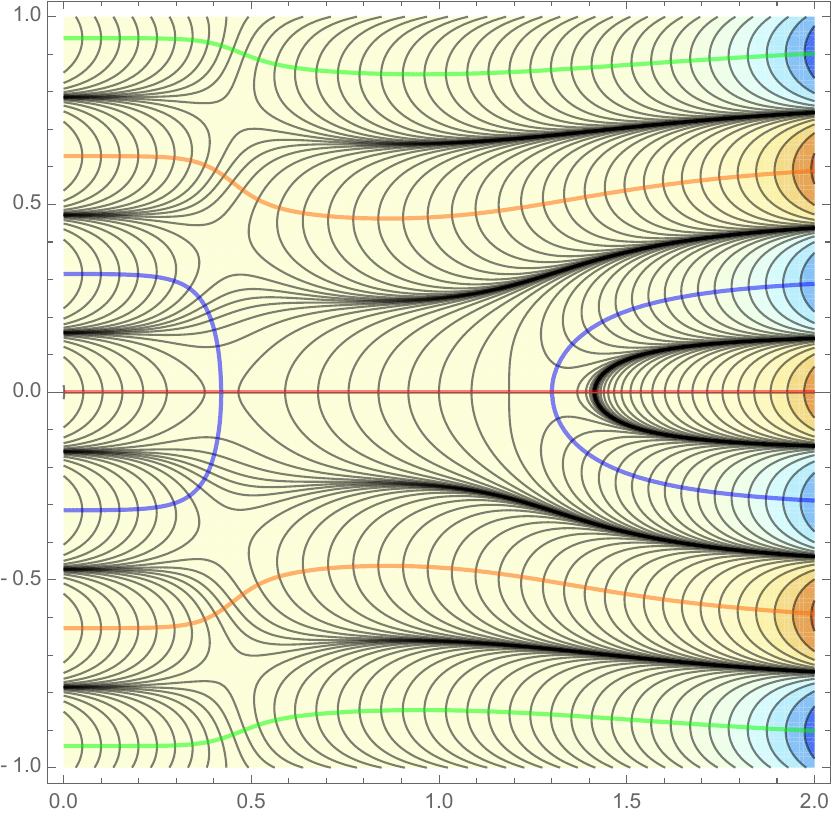}};
            \node[] at (1.3,4.375) {\large I};
            \node[] at (4,4.375) {\large II};
            \node[] at (7,4.375) {\large III};
        \end{tikzpicture}
    \caption{$\sinh x_1$}\label{fig:Tracks_sinh}
    \end{subfigure}
    \caption{The classical paths in the complex initial velocity plane for the initial position $x_0=-5$, barrier strength $V_0=1$, and propagation time $T=10$. The numbers I, II, and III, mark the direct and bouncing real classical paths.}
\end{figure}

This observation is unsurprising, as real classical paths are unlikely to capture the classically forbidden phenomenon of quantum tunneling. Equation \eqref{eq:PathIntegral2} suggests that this discrepancy follows from relevant complex paths. We can search for these complex paths by considering the initial value problem with a complex initial velocity. Consider the final position $f(v_0)= x_C(x_0,v_0;T)$ as a map from complex initial velocity space to complex final position space (see fig.\ \ref{fig:Tracks}). The reality condition, $\text{Im}[f(v_0)]=0$, defines a set of curves corresponding to the classical paths that terminate on the real line forming candidate classical paths. 

\begin{figure}
    \centering
    \begin{subfigure}[b]{0.32\textwidth}
        \includegraphics[width =\textwidth]{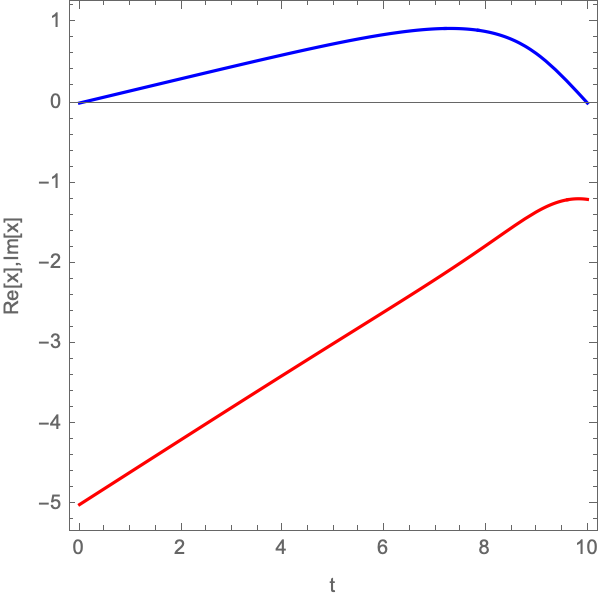}
    \end{subfigure}
    \begin{subfigure}[b]{0.32\textwidth}
        \includegraphics[width =\textwidth]{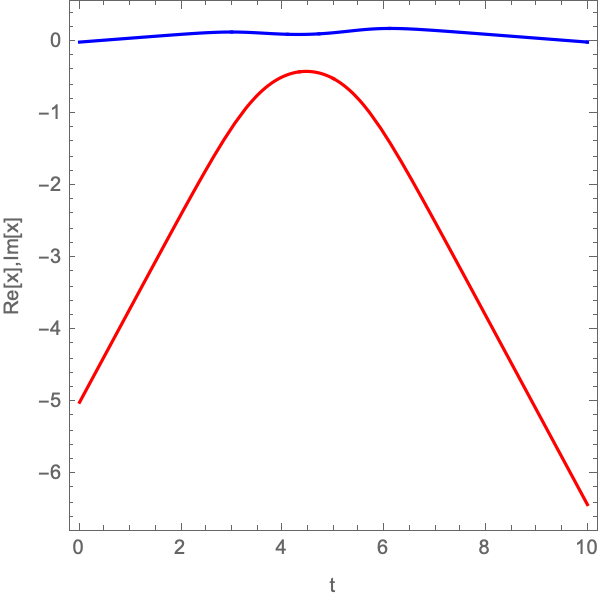}
    \end{subfigure}
    \begin{subfigure}[b]{0.32\textwidth}
        \includegraphics[width =\textwidth]{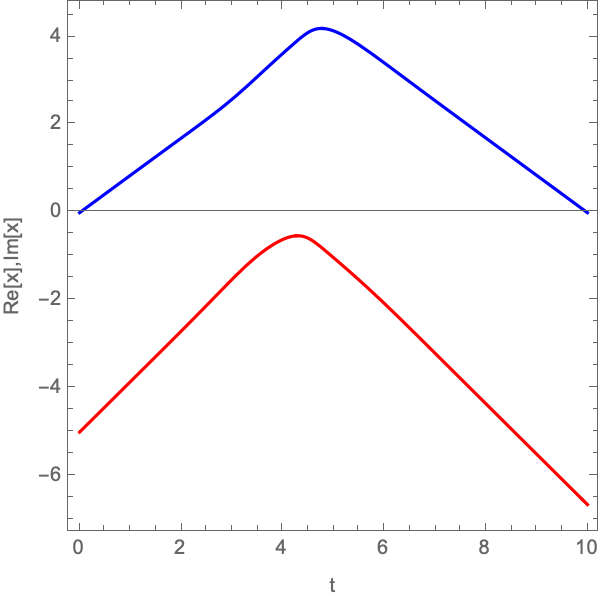}
    \end{subfigure}
    \begin{subfigure}[b]{0.32\textwidth}
        \includegraphics[width =\textwidth]{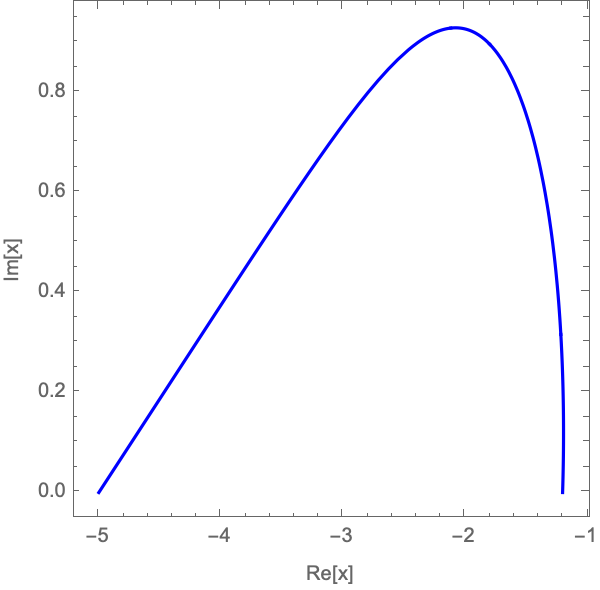}
    \end{subfigure}
    \begin{subfigure}[b]{0.32\textwidth}
        \includegraphics[width =\textwidth]{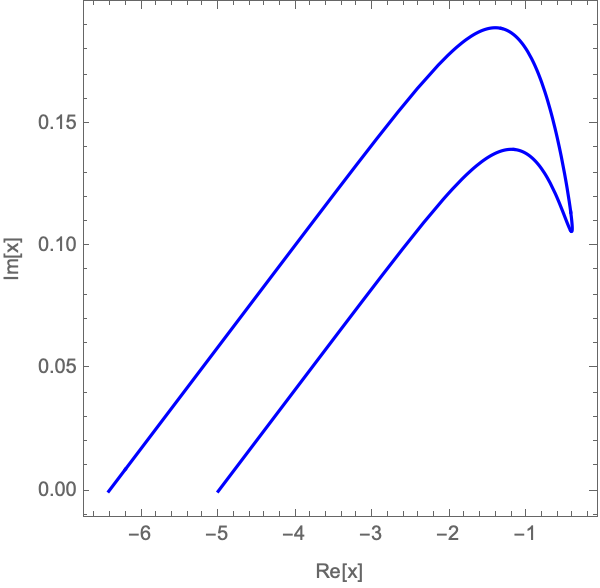}
    \end{subfigure}
    \begin{subfigure}[b]{0.32\textwidth}
        \includegraphics[width =\textwidth]{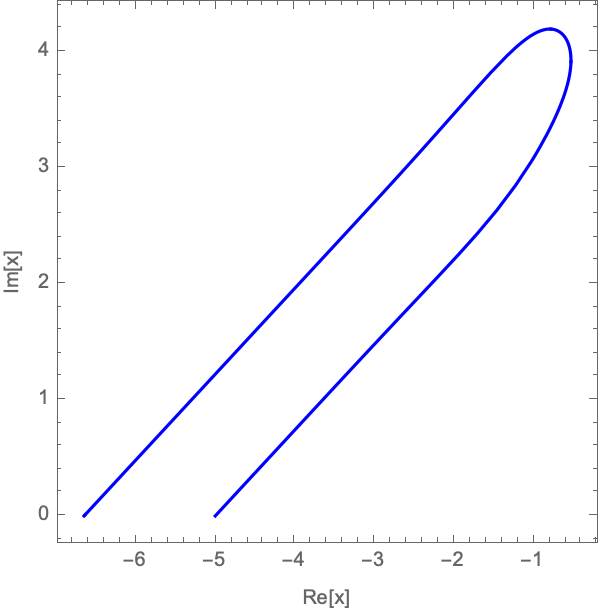}
    \end{subfigure}
    \caption{The complex paths as a function of time (top) and in the complex plane (bottom). \textit{Left:} a classical path on the left blue arc. \textit{Centre:} a classical path on the right blue arc. \textit{Right:} a classical path on the green arc.}
    \label{fig:complexPaths}
\end{figure}

The real line satisfies the reality condition as a real initial velocity always yields a real final position. We can identify the direct path, the bouncing solution with the lower initial velocity, and the bouncing solution that transitions into the classical path moving over the barrier with three segments of the real line. These three classes of real classical paths are represented by the red line segments labeled I, II, and III in fig.\ \ref{fig:Tracks}. Where these line segments meet, corresponding with the emergence or annihilation of two real classical paths, we find a branch of classical paths terminating at a real final position (the blue arcs). This is a characteristic property of a caustic. As we transition through a caustic, the merging real classical paths become complex conjugate paths. These complex paths $x_C$ form topological semicircles in complex position space (see fig.\ \ref{fig:complexPaths}).

Besides the families of complex classical paths associated with the caustics, we find an infinite set of families that do not intersect the real line as we vary $x_1$. These complex paths are also topological semi-circles ranging further into the complex position plane (see fig.\ \ref{fig:complexPaths}). Finally, observe that the blue and green arcs in fig.\ \ref{fig:Tracks} terminate when intersecting a set of white lines. As we vary the finial position $x_1$, the complex path can hit a singularity of the analytic continuation of the potential. The Rosen-Morse potential $V(x)=\cosh^{-2}(x)$ has a pole at $x = i \pi (n+1/2)$ for $n \in \mathbb{Z}$. In other words, for fixed values of $x_0$ and $T$, as we vary the final position, $x_1$, the resulting complex solution, $x_C(t)$, may coincide with a pole: $x_C(t) = i \pi (n + 1/2)$ for some $t$. After hitting the singularity, the classical path no longer terminates at a real final position, and, thereby, ceases to be a solution to the boundary-value problem. 

Using the analytic solution of the classical path \eqref{eq:sol1}, we can identify these singularity crossings with the branch cuts of the arcsinh function when insisting the classical path be continuous in time. Indeed, the change of variables $s_C(t)=\sinh(x_C(t))$ removes the singularity crossings (see fig.\ \ref{fig:Tracks_sinh}). However, note that this transformation is not one-to-one in the complex plane as $\sinh(i n \pi) = 0$ for all $n \in \mathbb{Z}$. Even though the family of complex classical paths terminates at the singularity crossing, we can analytically continue the final position, $f(v_0)$, beyond this barrier when interpreted as a branch cut. The analytic continuation satisfying the reality condition is indicated by the dashed black curves. Note that on the black curve, the initial velocity $v_0$ serves only as a label of the classical path, as the associated classical action does not solve the boundary value problem. Instead, the path terminates at $(-1)^n x_1 \pm n \pi i$ for some nonzero integer $n$.

\begin{figure}
    \centering
    \begin{subfigure}[b]{0.49\textwidth}
        \includegraphics[width =\textwidth]{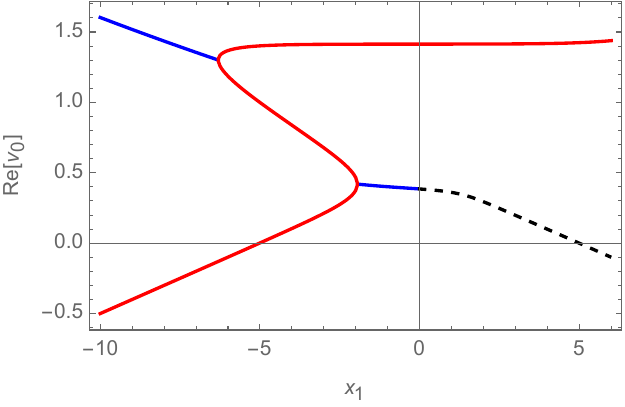}
    \end{subfigure}
    \begin{subfigure}[b]{0.49\textwidth}
        \includegraphics[width =\textwidth]{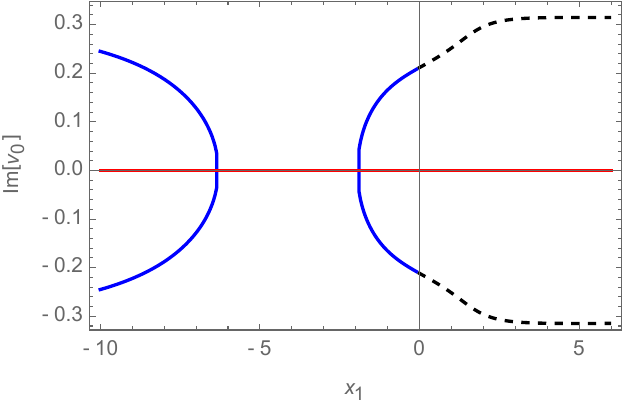}
    \end{subfigure}\\
    \begin{subfigure}[b]{0.49\textwidth}
        \includegraphics[width =\textwidth]{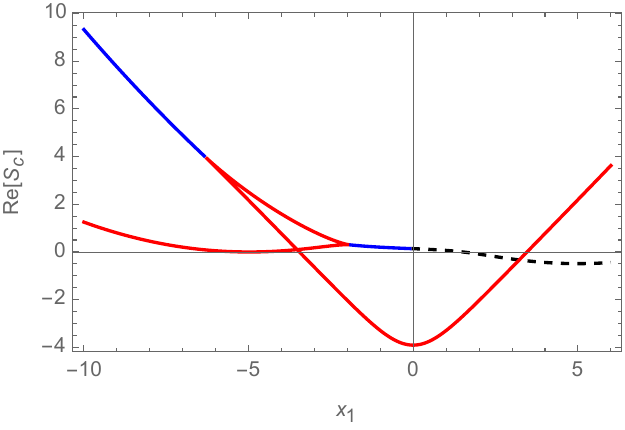}
    \end{subfigure}
    \begin{subfigure}[b]{0.49\textwidth}
        \includegraphics[width =\textwidth]{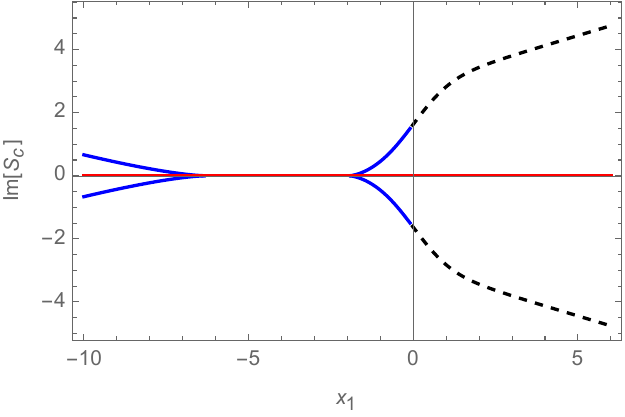}
    \end{subfigure}
    \caption{The initial velocity and action as a function of the final position $x_1$ for initial position $x_0=-5$, mass $m=1$, time $T=10$ and barrier strength $V_0=1$. The red curves correspond to the real classical paths. The blue curves correspond to the complex classical paths. The dashed black curves correspond to the analytically continued complex classical paths. The complex classical paths with a positive imaginary part of the action are relevant while the one with a negative imaginary part of the action is irrelevant to the path integral.}
    \label{fig:complex_velocity_action}
\end{figure}

The complex conjugate pair of complex paths associated with the blue arcs form a natural continuation of the real classical paths annihilating at a caustic (see fig.\ \ref{fig:complex_velocity_action}). According to Picard-Lefschetz theory, the complex path, with $f(v_0)=x_1$, for which the real part of the exponent $iS_C$ is negative is generally relevant near the caustic. The conjugate path, with $\text{Re}[iS_C] >0$ is irrelevant to the path integral. 

At the singularity crossing, appearing when $x_1$ approaches a small negative value from below, the complex path ceases to exist. The naive saddle point approximation fails to capture real-time quantum tunneling as there is no associated complex solution to the classical boundary value problem. Instead at the singularity crossing of the Rosen-Morse theory, the endpoint transforms as $x_1 \mapsto - x_1 \pm i \pi$. However, the analytic continuation of the final position $f(v_0)$ and the classical action $s(v_0) = S[x_C(x_0,v_0;T)]$ beyond the singularity crossing yields a solution. Solving $f(v_0)=x_1$ beyond the branch cut we find a complex initial velocity serving only as a label. In particular, the initial velocity $v_0(x_1)$ follows a continuous path along the blue arc and subsequently along the dashed black extension while continuously increasing $x_1$. The classical action $s(v_0) = S[x_C(x_0,v_0;T)]$ is also discontinuous at the singularity crossing, acquiring a shift of $\pm i \sqrt{2mV_0}\pi$, and can be analytically continued like the final position $f(v_0)$. When evaluating the function $s$ along the path $v_0(x_1)$ we obtain an analytically continued action (see the dashed black curves in fig.\ \ref{fig:complex_velocity_action}) that can be used to extend the traditional saddle point approximation. Note that the initial velocity now only serves as a label, as there no longer exists a continuously deformed complex classical path solving the boundary value problem \footnote{In the $s_C$ variable, the associated final position $\sinh(f(v_0))$ is free of branch cuts. However, beyond the white curves, the associated classical path $x_C$ doesn't solve the boundary value problem. The classical action $S_C$ has discontinuities regardless of whether we use the $x_C$ of the $s_C$ variables.}.

\begin{figure}
    \centering
    \begin{subfigure}[b]{0.49\textwidth}
        \includegraphics[width =\textwidth]{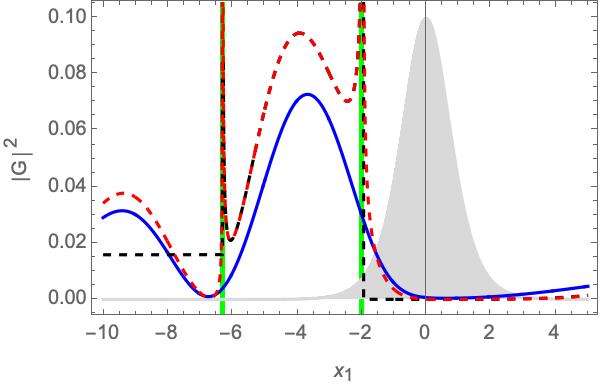}
        \caption{$\hbar=1$}
    \end{subfigure}
    \begin{subfigure}[b]{0.49\textwidth}
        \includegraphics[width =\textwidth]{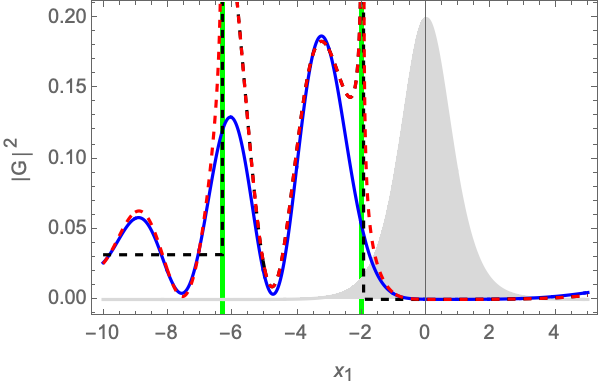}
        \caption{$\hbar=0.5$}
    \end{subfigure}\\
    \begin{subfigure}[b]{0.49\textwidth}
        \includegraphics[width =\textwidth]{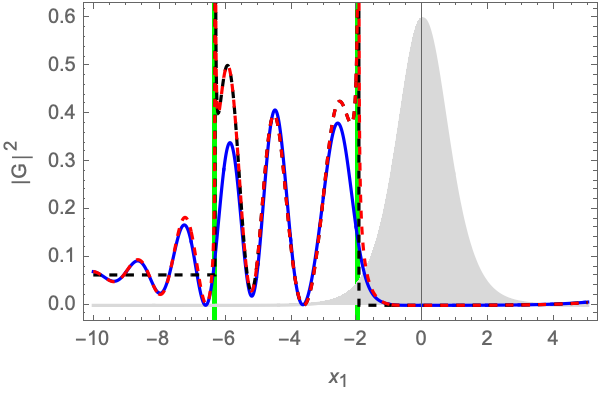}
        \caption{$\hbar=0.25$}
    \end{subfigure}
    \begin{subfigure}[b]{0.49\textwidth}
        \includegraphics[width =\textwidth]{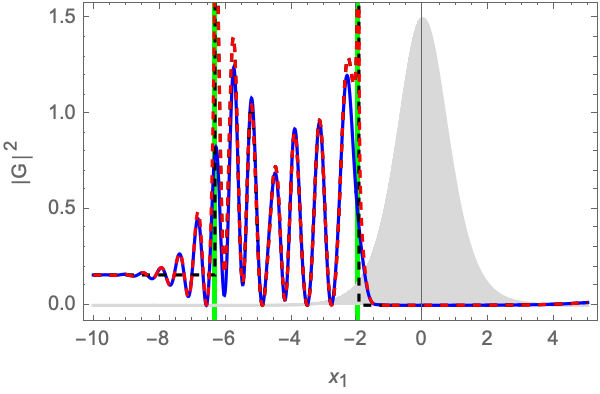}
        \caption{$\hbar=0.1$}
    \end{subfigure}
    \caption{The Feynman propagator squared as a function of the final position $x_1$ for the initial position $x_0=-5$, mass $m=1$ time $T=10,$ and barrier strength $V_0=1$ for $\hbar=0.1, 0.25,0.5,1$.  The exact (blue), the saddle point approximation based on the real classical paths (black) and the one based on the relevant classical paths (red). The caustics are marked by green vertical lines.}
    \label{fig:WKB_line}
\end{figure}

Given the analysis above, we can extend the saddle point approximation to include the relevant complex path associated with the caustic (see fig.\ \ref{fig:WKB_line}). Inside the caustic region, the two saddle point approximations (based on the real and the relevant (complex) classical paths) coincide. The resulting approximation neatly captures the transition from the region with three to one real classical paths. The complex paths associated with the green arcs, and the other families disjoint from the caustics, turn out to be irrelevant as they lead to large discrepancies with respect to the exact propagator in the saddle point approximation. The real paths and complex paths associated with the caustics (and their analytic continuation) are the relevant classical paths and suffice to define the path integral.

\begin{figure}
    \centering
    \begin{tikzpicture}
        \node[anchor=south west,inner sep=0] at (0,0) {\includegraphics[width = 0.7\textwidth]{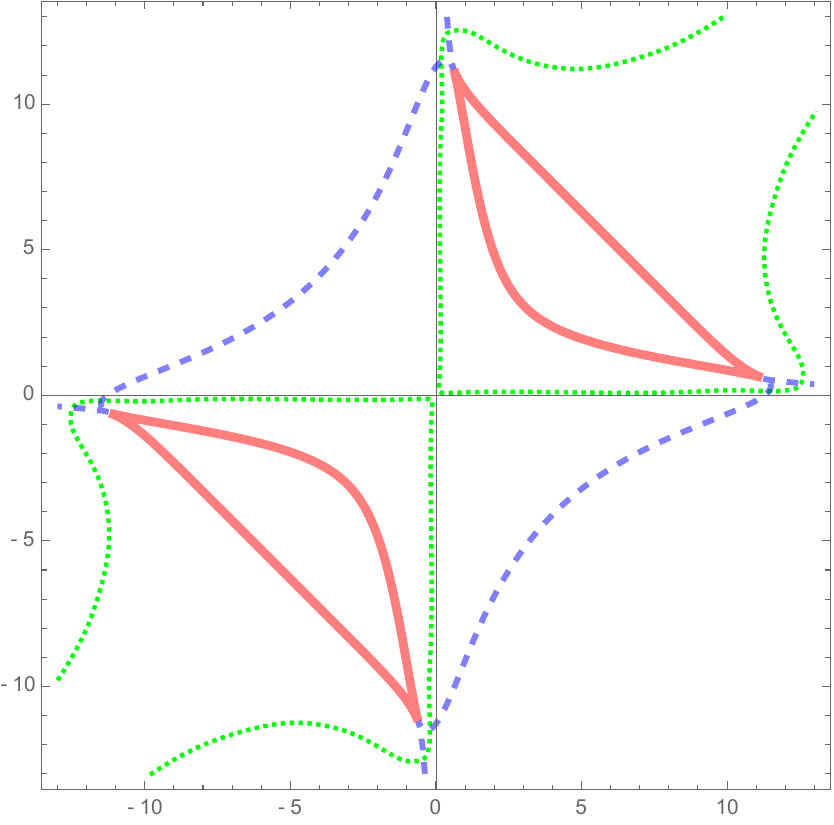}};
        \node[draw,circle,minimum size=0.8cm,inner sep=0pt] at (4.25,4.25) {\Large I};
        \node[draw,circle,minimum size=0.8cm,inner sep=0pt] at (7.75,7.75) {\Large I};
        \node[draw,circle,minimum size=0.8cm,inner sep=0pt] at (3,3) {\Large II};
        \node[draw,circle,minimum size=0.8cm,inner sep=0pt] at (9,9) {\Large II};
        \node[draw,circle,minimum size=0.8cm,inner sep=0pt] at (5.25,5.25) {\Large III};
        \node[draw,circle,minimum size=0.8cm,inner sep=0pt] at (6.75,6.75) {\Large III};
        \node[draw,circle,minimum size=0.8cm,inner sep=0pt] at (6.75,5.25) {\Large IV};
        \node[draw,circle,minimum size=0.8cm,inner sep=0pt] at (5.25,6.75) {\Large IV};
        \node[draw,circle,minimum size=0.8cm,inner sep=0pt] at (9,3) {\Large VI};
        \node[draw,circle,minimum size=0.8cm,inner sep=0pt] at (3,9) {\Large VI};
        \node[draw,circle,minimum size=0.8cm,inner sep=0pt] at (11,7.75) {\Large V};
        \node[draw,circle,minimum size=0.8cm,inner sep=0pt] at (7.75,11) {\Large V};
        \node[draw,circle,minimum size=0.8cm,inner sep=0pt] at (1,4.25) {\Large V};
        \node[draw,circle,minimum size=0.8cm,inner sep=0pt] at (4.25,1) {\Large V};
    \end{tikzpicture}
    \caption{The caustic curves (solid red), the Stoke's curves (dashed blue), and the singularity crossing curves (dotted green) in the $(x_0,x_1)$-plane of the Teller potential for the mass $m=1$, barrier strength $V_0=1$ and propagation time $T=10$.}\label{fig:Caustics+Stokes}
\end{figure}

The analysis above is restricted to the line $x_0=-5$. When considering the space of initial and final positions $(x_0,x_1)$, the relevant classical paths are governed by the caustic, singularity crossing, and Stoke's curves (see fig.\ \ref{fig:Caustics+Stokes}). At the caustics (the red curves), classical paths merge and transition from real to complex. Generally, one of the complex paths is relevant after the transition. At singularity crossing curves, the classical path ceases to exist as a solution to the boundary value problem. At Stoke's curves, for which the imaginary part of the exponent $iS_C$ of two classical paths coincide, the relevant complex path becomes irrelevant \cite{Feldbrugge:2023AnPhy}.

\begin{figure}
    \centering
    \begin{subfigure}[b]{0.45\textwidth}
        \includegraphics[width =\textwidth]{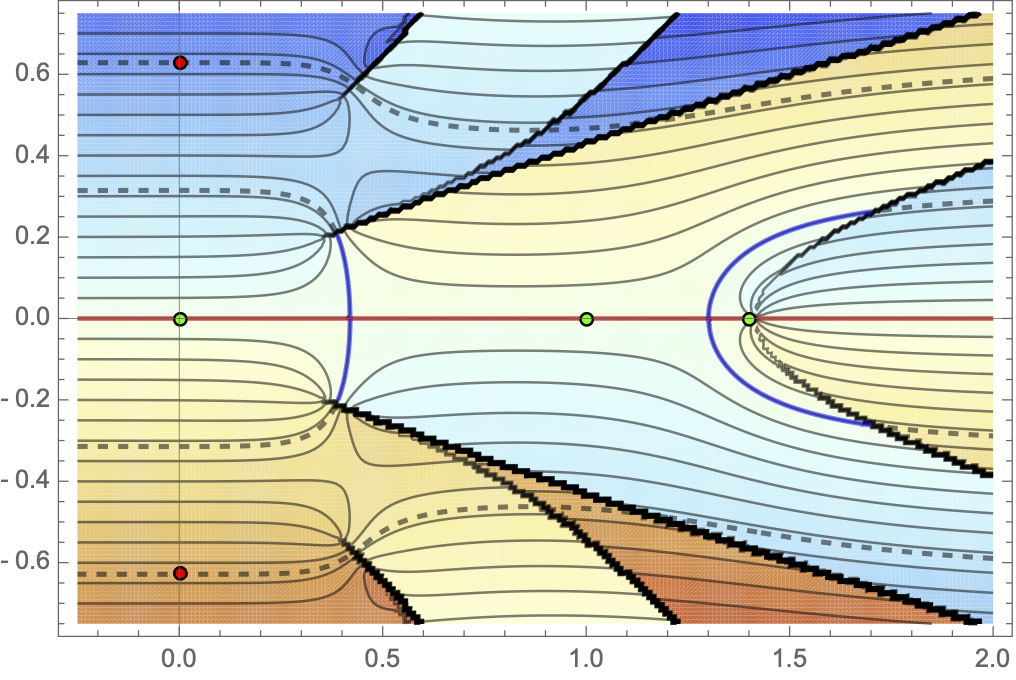}
        \caption{Region I}
    \end{subfigure}
    \begin{subfigure}[b]{0.45\textwidth}
        \includegraphics[width =\textwidth]{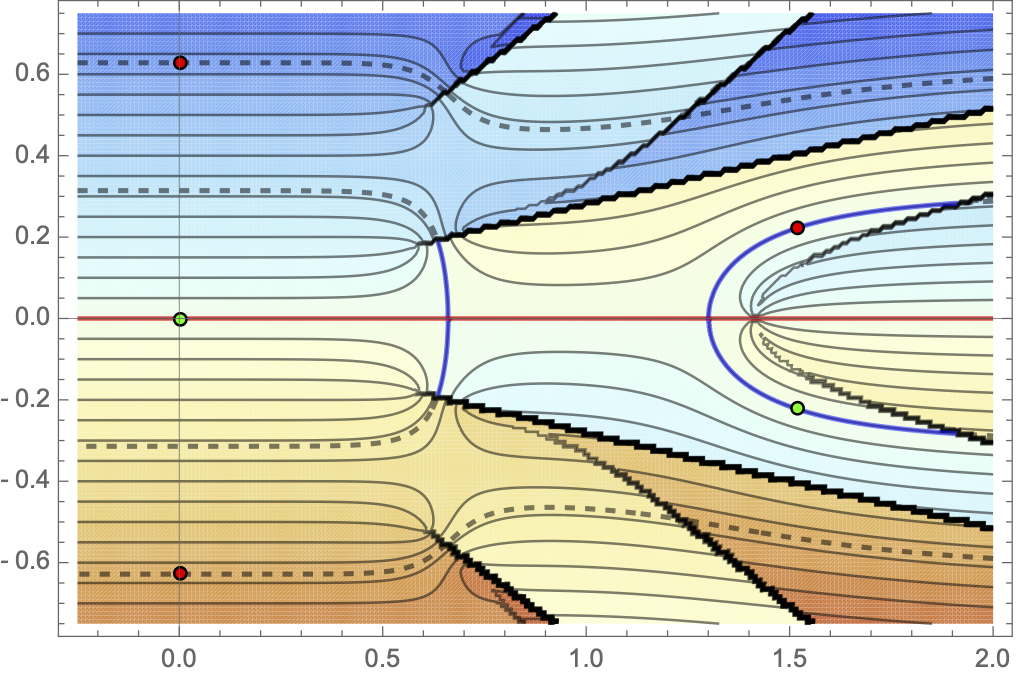}
        \caption{Region II}
    \end{subfigure}\\
    \begin{subfigure}[b]{0.45\textwidth}
        \includegraphics[width =\textwidth]{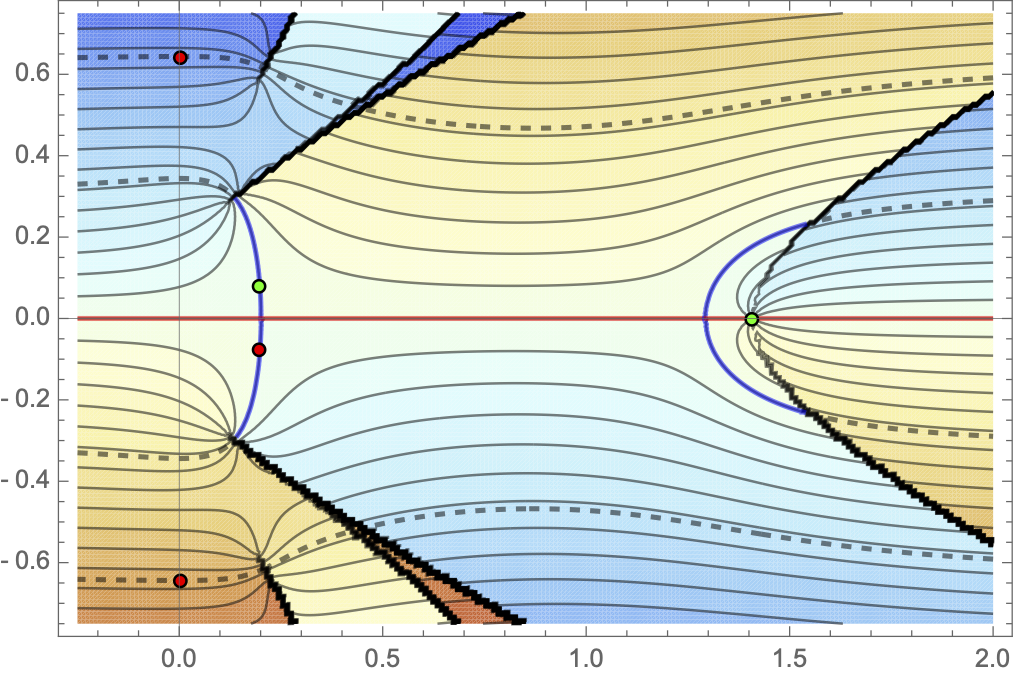}
        \caption{Region III}
    \end{subfigure}
    \begin{subfigure}[b]{0.45\textwidth}
        \includegraphics[width =\textwidth]{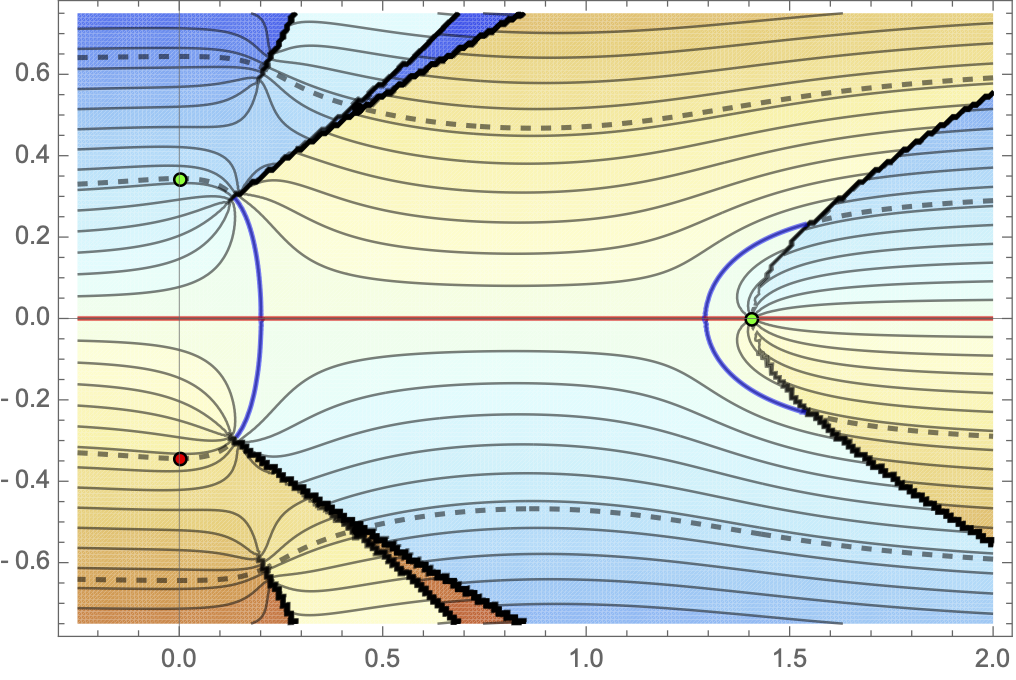}
        \caption{Region IV}
    \end{subfigure}\\
    \begin{subfigure}[b]{0.45\textwidth}
        \includegraphics[width =\textwidth]{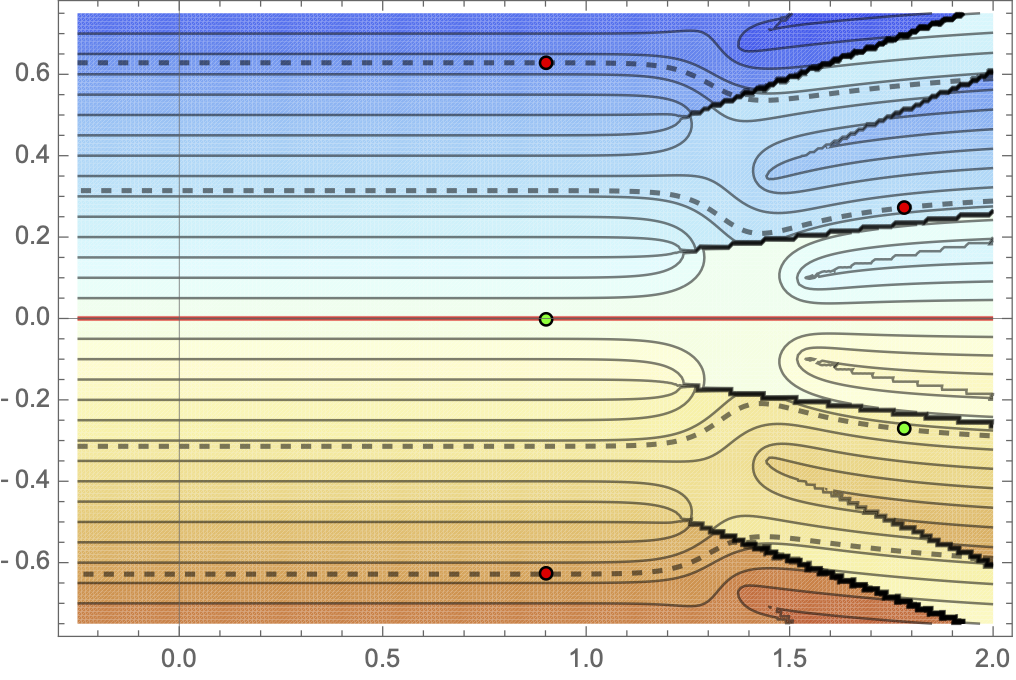}
        \caption{Region V}
    \end{subfigure}
    \begin{subfigure}[b]{0.45\textwidth}
        \includegraphics[width =\textwidth]{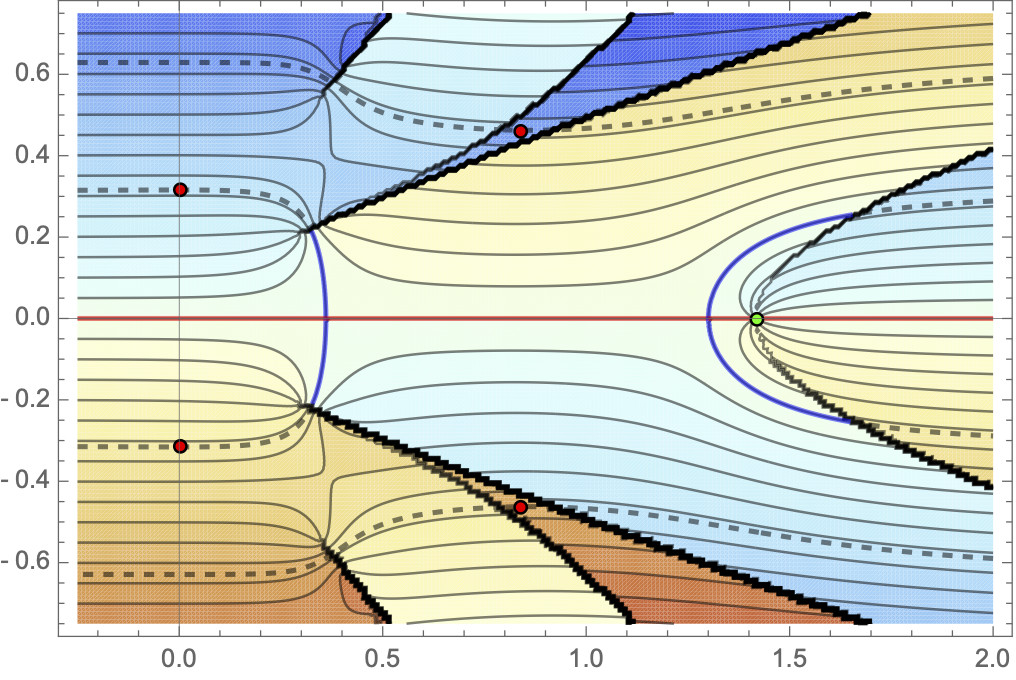}
        \caption{Region VI}
    \end{subfigure}\\
    \caption{The classical paths in the six regions listed in fig.\ \ref{fig:Caustics+Stokes} in the complex initial velocity plane. Relevant classical paths are plotted in green. Irrelevant classical paths are plotted in red. The horizontal red line on the real axis represents real classical paths. The blue arcs represent complex solutions to the boundary value problem. The black dashed lines represent analytically continued classical paths.}
    \label{fig:regions}
\end{figure}

We identify six qualitatively different regions. Region $1$ has three real relevant classical paths. Region $2$ and region $3$ has one real and one complex relevant classical path. The complex classical path lies on either the left or the right blue arc. In regions $4$ and $5$ we have one real and one analytically continued complex relevant path. Finally, in region $6$ we find only a single real relevant classical path. We identify the relevant classical paths in fig.\ \ref{fig:regions}. Note that at the cusp caustic the two blue arcs merge. For $x_0$ smaller than the initial position corresponding to the cusp caustic the blue arcs no longer intersect the real axis. 

%%%%%%%%%%%%%%%%%%%%%%%%%%
\subsection{Evolution of wavefunctions}
In the previous section, we analyzed the role of the (complex) classical solutions of the associated boundary value problem in the Feynman path integral. Moreover, we saw how caustics can describe the characteristic behavior of a reflecting wavefunction (see fig.\ \ref{fig:Schrodinger}). In this section, we refine this analysis by studying the implications of real and complex classical paths in wave functions in the semi-classical limit.

The Schr\"odinger equation is solved by a convolution of the initial wavefunction with the Feynman propagator. Let's consider the coherent initial state
\begin{align}
    \psi_0(x_0) = \frac{1}{\sqrt[4]{2 \pi \sigma_0^2}} e^{-\frac{(x_0-\bar{x}_0)^2}{4 \sigma_0^2} + i \frac{\bar{p}_0 x_0}{\hbar}}\,.
\end{align}
In the semi-classical limit, the convolution assumes the form
\begin{align}
    \psi_T(x_1) &= \int G[x_1,x_0;T]\psi_0(x_0)\mathrm{d}x_0 \\
    &\sim \Theta_H(T) \sqrt{\frac{i}{2\pi \hbar}} \frac{1}{\sqrt[4]{2 \pi \sigma_0^2}}\sum_{x_C} \sqrt{\frac{\partial^2 S_C}{\partial x_0 \partial x_1}}  \int e^{-\frac{(x_0-\bar{x}_0)^2}{4 \sigma_0^2} + i \frac{S_C + \bar{p}_0 x_0}{\hbar} }\mathrm{d}x_0\\
    &\sim \Theta_H(T)\sqrt{\frac{i}{2\pi \hbar}} \frac{1}{\sqrt[4]{2 \pi \sigma_0^2}}\sum_{x_C} \sqrt{\frac{\partial^2 S_C}{\partial x_0 \partial x_1}}  \int e^{-\frac{(x_0-\bar{x}_0)^2}{4 \sigma_0^2} + i \frac{\bar{S}_C + \partial_{x_0} \bar{S}_C (x_0-\bar{x}_0) + \dots+\bar{p}_0 x_0}{\hbar}}\mathrm{d}x_0\\
    &\sim \Theta_H(T)\sqrt[4]{\frac{-2 \sigma_0^2 }{\pi \hbar^2}} \sum_{x_C} \sqrt{\frac{\partial^2 S_C}{\partial x_0 \partial x_1}}  e^{-\frac{(\bar{p}_0 + \partial_{x_0}\bar{S}_C)^2 \sigma_0^2}{\hbar^2} + i \frac{\bar{p}_0 \bar{x}_0}{\hbar}} e^{i \bar{S}_C/\hbar}\,,
\end{align}
upon expanding the action $S_C$ around $\bar{x}_0$ to linear order, \textit{i.e.}, 
\begin{align}
    \bar{S}_C = \bar{S}_C + \partial_{x_0} \bar{S}_C (x_0 - \bar{x}_0) + \mathcal{O}\left((x_0-\bar{x}_0)^2\right)\,,
\end{align}
summing over the relevant classical paths associated with the pair $(x_0,x_1)$. Upon the convolution with the Gaussian initial state, each relevant classical path receives a weight
\begin{align}
    w_C = \exp\left[-\frac{(\bar{p}_0 + \partial_{x_0}\bar{S}_C)^2 \sigma_0^2}{\hbar^2} + i\frac{(\bar{S}_C +\bar{p}_0 \bar{x}_0)}{\hbar}\right]\,,
\end{align}
amplifying classical paths for which the initial momentum is close to the mean momentum of the initial state $\bar{p}_0$, as by the Hamilton-Jacobi equation, 
\begin{align}
    p_{0,C} = m v_{0,C}= -\partial_{x_0} \bar{S}_C \,,
\end{align}
with the initial momentum $p_{0,C}$ and initial velocity $v_{0,C}$ of the classical path $x_C$. Classical paths with initial momenta separated from $\bar{p}_0$ are suppressed.

Note that this equation holds for both real and complex classical paths. Indeed, the norm of the weight yields the Gaussian factor
\begin{align}
    |w_C| = \exp\left[-\frac{(\bar{p}_0 - m\, \text{Re}[v_{0,C}])^2 \sigma_{0}^2}{\hbar^2} + \frac{m^2\, \text{Im}[v_{0,C}]^2 \sigma_0^2}{\hbar^2} - \frac{\text{Im}[\bar{S}_C]}{\hbar}\right]\,.
\end{align}
with $\text{Re}[v_{0,C}]$ and $\text{Im}[v_{0,C}]$ the real and imaginary part of the initial velocty. For real classical paths, the weight, $|w_C|$, of the path with initial momentum $p_{0,C}$ is a Gaussian centered at $\bar{p}_0$. For complex classical paths, the weight is a Gaussian where the real part of the initial momentum is centered at $\bar{p}_0$. The imaginary part of the initial momentum enhances the weight, while the imaginary part of the corresponding classical action suppresses the weight. When working with analytically continued complex classical paths, the same logic applies when defining the effective initial momentum using the Hamilton-Jacobi equation $p_{0,C} = - \partial_{x_0}\bar{S}_C$

Given a general initial state, $\psi_0$, we can expand the state in terms of coherent states and find the significance of the relevant classical paths in the semi-classical limit using these weights. For example, in fig.\ \ref{fig:Schrodinger}, either a single or three real classical paths significantly influence the final wavefunction. In the lower-left panel, we retrieve the dominant real classical path. In the lower-right panel, we observe the interference pattern resulting from the three real classical paths. Conversely, the relevance of (complex) classical paths can be probed by convolving the exact Green's function with various coherent states.

%%%%%%%%%%%%%%%%%%%%%%%%%%
\section{The energy propagator}\label{sec:energy}
In the previous sections, we looked into the structure of the real-time Feynman path integral. To investigate the nature of quantum reflection and tunneling more closely we turn toward the energy propagator, which colloquially assumes the form
\begin{align}
    K[x_1,x_0;E]\, ``\hspace{-1mm}=\hspace{-1mm}" \int_0^\infty \int_{x(0)=x_0}^{x(T)=x_1} e^{i (S[x] + E T ) /\hbar}\mathcal{D}x\, \mathrm{d}T\,.
\end{align}
Using the Picard-Lefschetz theory, we propose the more formal integral
\begin{align}
    K[x_1,x_0;E] = \sqrt{\frac{m}{2E}} \sum_{(x_C,T_C)} e^{i(S_C+E T_C)/\hbar} \int_{\mathcal{J}_C} e^{i \theta_C} \mathrm{d} \mu_C\,,
\end{align}
where the sum ranges over the relevant classical paths and stationary points of the time integral pairs, satisfying the variational equations
\begin{align}
    \frac{\delta S}{\delta x} = 0\,,\quad E+ \frac{\partial S}{\partial T}=0\,,
\end{align}
and the boundary conditions $x(0)=x_0$ and $x(T)=x_1$. Using the saddle point approximation, we obtain the approximate propagator
\begin{align}
    K_{WKB}[x_1,x_0;E] &= \sum_{(x_C,T_C)} \sqrt{-\frac{\partial^2 S_C/\partial x_0 \partial x_1}{\partial^2S_C/\partial T^2}} e^{i(S_C+ET_C)/\hbar}\\
    &= \sum_{(x_C,T_C)} \sqrt{-\frac{\partial T_C/\partial E}{\partial x_1/\partial p_0}} e^{i(S_C+ET_C)/\hbar}\,.
\end{align}
where the functional determinant includes an additional contribution from the time integral (see \cite{Schulman:2012}) which we write using the Hamilton-Jacobi equation $\partial S_C/\partial x_0= - p_0$. For an illustrative example of the saddle point approximation of the energy propagator, we summarize the propagator of a particle in a linear potential in appendix \ref{ap:Linear}.

To evaluate the stationary points of the energy propagator, allowing for complex integration times $T$, we rescale the time parameter $t = T \lambda$ and write the action as
\begin{align}
    S[x] = \int_0^1 \left[\frac{m x'(\lambda)^2}{2 T} - \frac{V_0 T}{\cosh^2(x(\lambda))} \right]\mathrm{d}\lambda\,,
\end{align}
with the prime denoting differentiation with respect to $\lambda$, yielding the equations of motion
\begin{align}
    \frac{m x''}{T^2} = \frac{2 V_0 \tanh x}{\cosh^2 x}\,, \quad
    E = \int_0^1 \left[ \frac{m x'(\lambda)^2}{2T^2} + \frac{V_0}{\cosh^2 x(\lambda)}\right]\mathrm{d}\lambda =
    \frac{m x'(0)^2}{2T^2} + \frac{V_0}{\cosh^2 x_0}\,,
\end{align}
where the initial velocity $v_0 = \dot{x}(0)= x'(0)/T$. The associated initial value problem
\begin{align}
    x_C(0)=x_0\,, \quad x_C'(0) = \pm \sqrt{\frac{2\left(E-V_0\, \sech^2 x_0\right)}{m}}T
\end{align}
is solved by the identity
\begin{align}
    \sinh(x_C(\lambda)) = c_1 \sqrt{\frac{E-V_0}{E}} \sinh\left[ \sqrt{\frac{2E}{m}}(T \lambda -C) \right]\,,
\end{align}
with the shift parameter
\begin{align}
    C = c_2 \sqrt{\frac{m}{2E}} \sinh^{-1}\left[ \sqrt{\frac{E}{E-V_0}} \sinh(x_0) \right]\,,
\end{align}
where we have to select the signs $c_1,c_2 = \pm 1$ the match the initial conditions.

%%%%%%%%%%%%%%%%%%%%%%%%%%
\subsection{The high energy propagator}
\begin{figure}
    \centering
    \begin{subfigure}[b]{0.49\textwidth}
        \includegraphics[width =\textwidth]{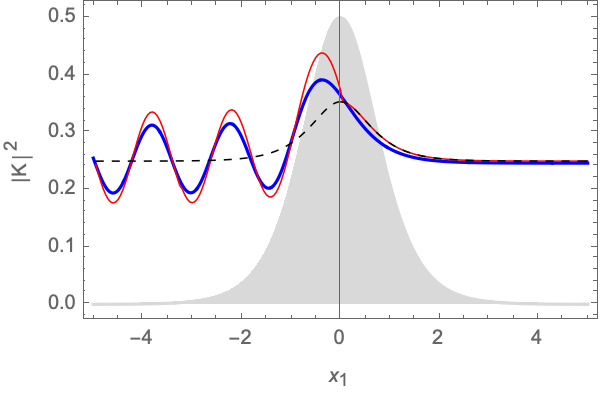}
        \caption{$E=2, \hbar=1$}
    \end{subfigure}
    \begin{subfigure}[b]{0.49\textwidth}
        \includegraphics[width =\textwidth]{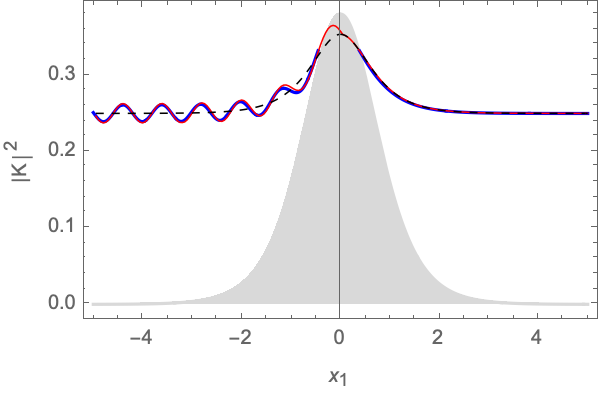}
        \caption{$E=2, \hbar=0.5$}
    \end{subfigure}\\
    \begin{subfigure}[b]{0.49\textwidth}
        \includegraphics[width =\textwidth]{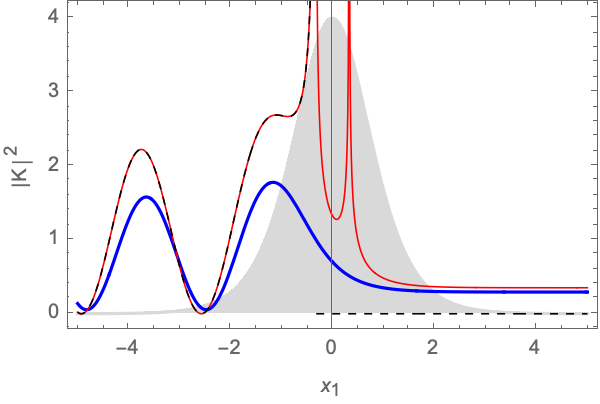}
        \caption{$E=0.9, \hbar=1$}
    \end{subfigure}
    \begin{subfigure}[b]{0.49\textwidth}
        \includegraphics[width =\textwidth]{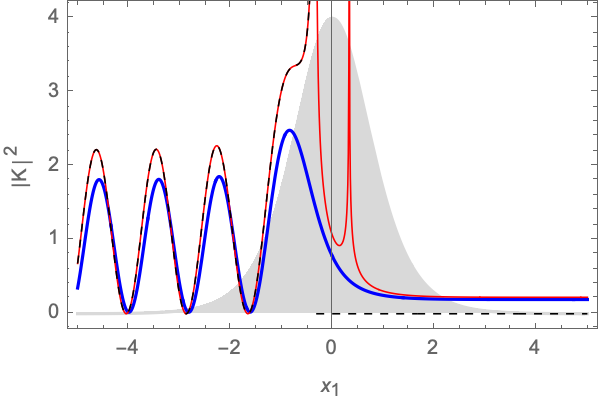}
        \caption{$E=0.9, \hbar=0.5$}
    \end{subfigure}
    \caption{The energy propagator squared as a function of the final position $x_1$ for the initial position $x_0=-5$, mass $m=1$, propagation time $E=2,$ and barrier strength $V_0=1$ for $\hbar=0.5,1$.  The exact (the blue curve), the saddle point approximation based on the real classical paths (the dashed black curve), and the one based on the relevant classical paths (the red curve).}
    \label{fig:High_E}
\end{figure}

When the energy $E$ exceeds the top of the potential $V_0$, any transition is classically allowed. The particle is unable to classically bounce off the barrier and the boundary value problem is solved by a single real classical path. However, depending on the boundary conditions, the real classical path is not the only relevant path to the propagator. The real path works well for paths moving over the barrier but fails for paths starting and ending on the same side of the barrier (see the upper panels of fig.\ \ref{fig:High_E}). 

\begin{figure}
    \centering
    \begin{subfigure}[b]{0.49\textwidth}
        \includegraphics[width =\textwidth]{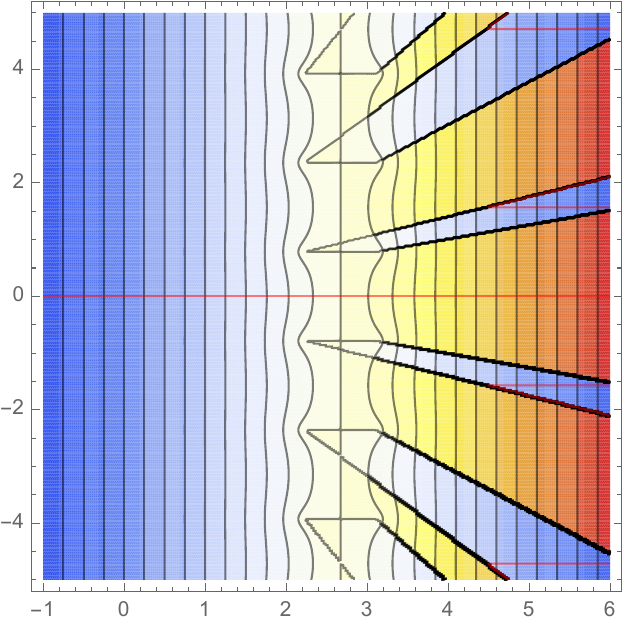}
        \caption{$E = 2, x_1$}
    \end{subfigure}
    \begin{subfigure}[b]{0.49\textwidth}
        \includegraphics[width =\textwidth]{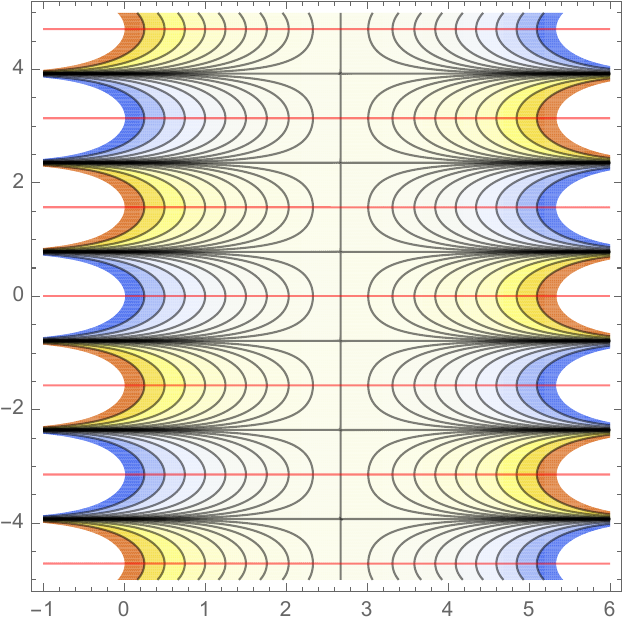}
        \caption{$E = 2, \sinh x_1$}
    \end{subfigure}\\
    \begin{subfigure}[b]{0.49\textwidth}
        \includegraphics[width =\textwidth]{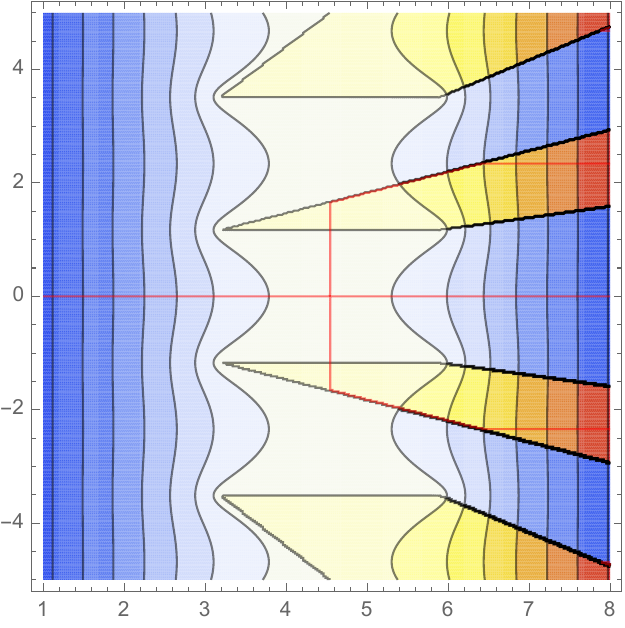}
        \caption{$E = 0.9, x_1$}
    \end{subfigure}
    \begin{subfigure}[b]{0.49\textwidth}
        \includegraphics[width =\textwidth]{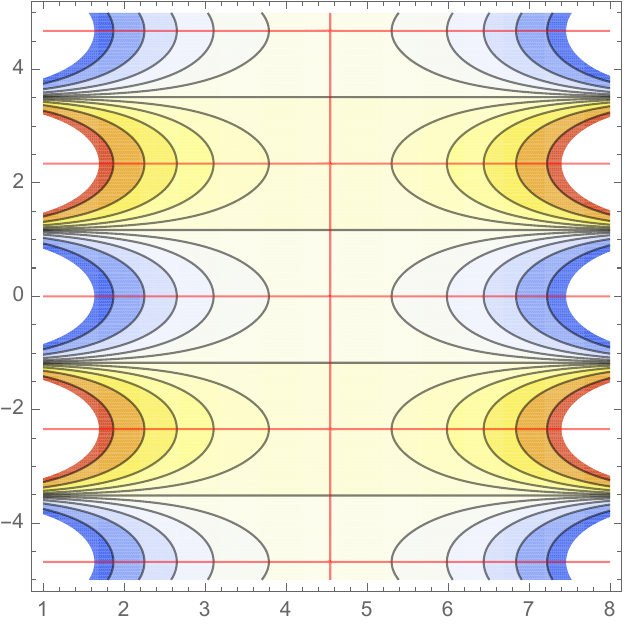}
        \caption{$E = 0.9, \sinh x_1$}
    \end{subfigure}
    \caption{The classical paths in the complex $T$ plane for the initial position $x_0=-5$, the mass $m=1$, barrier strength $V_0=1$, and energy $E=0.9,2$.}\label{fig:Tracks_High}
\end{figure}

The particle reflects off the barrier, even though this is forbidden in the classical theory. This behavior can be explained in terms of a relevant complex classical path. Mirroring the discussion of the real-time propagator, we solve the initial value problem in the complex $T$ plane and evaluate the final position at $\lambda=1$ (see the upper panels of fig.\ \ref{fig:Tracks_High}). The real classical paths again reside on the real $T$ line. Note that $x_1$ is a monotonic function of the time $T$ (as there are no real bouncing solutions). In addition, the boundary value problem has a set of complex solutions corresponding to the horizontal lines parallel the the real axis. These families of complex paths undergo a singularity crossing as we increase $x_1$. The first complex path (on the blue line) captures the oscillations of the exact propagator when $x_0$ and $x_1$ reside on the same side of the barrier. When one of the boundary conditions crosses the top of the barrier at $x=0$ we observe a Stoke's phenomenon making the complex path irrelevant to the propagator (see the upper panels of fig.\ \ref{fig:High_E}).

%%%%%%%%%%%%%%%%%%%%%%%%%%
\subsection{The low energy propagator}
When the energy $E$ is lower than the height of the potential $V_0$, the behavior changes dramatically. We identify two classically allowed regions 
\begin{align}
    \left(-\infty, -\cosh^{-1}\sqrt{V_0/E}\right],\quad \left[\cosh^{-1}\sqrt{V_0/E}, \infty\right)
\end{align} 
and a classically forbidden region 
\begin{align}
    \left(-\cosh^{-1}\sqrt{V_0/E}, \cosh^{-1} \sqrt{V_0/E}\right)\,.
\end{align}
When the boundary conditions $(x_0,x_1)$ lay in the same classically allowed region, there exist two real classical paths, leading to an accurate saddle point approximation of the energy propagator. When the initial and final conditions lay on different sides of the barrier, there is no such real path. The real saddle point approximation fails in this regime (see the lower panels of fig.\ \ref{fig:High_E}). Starting with two boundary points $|x_0|,|x_1| > \cosh^{-1}\sqrt{V_0/E}$, the two real classical paths merge in a fold caustic as we let one of the two boundary conditions approach the turning point $x=\pm \cosh^{-1}\sqrt{V_0/E}$. After passing the turning point, the real classical paths form a complex conjugate pair of complex paths, one of which is relevant to the energy propagator (see the lower panels of fig.\ \ref{fig:High_E}). As we let the boundary value move through the classically forbidden region, it crosses a singularity crossing, at which we need to analytically continue the final position and classical action for the saddle point approximation. In the analytic continuation, we identify a complex caustic, where two complex classical paths merge at the intersections of the real lines in the lower right panel of fig.\ \ref{fig:Tracks_High}. This complex caustic corresponds to $x_1$ moving into the classically allowed region on the right of the barrier. After this caustic, we obtain one relevant complex classical path that turns to the left in the complex $T$ plane. Note that the energy propagator approaches a nonzero value when $x_0 \to -\infty$ and $x_1 \to \infty$ correspond to the quantum tunneling rate.

When changing the energy $E$ passed the height of the barrier $V_0$, the structure of the relevant complex paths changes dramatically in a caustic. We thus find that the saddle point approximation is not accurate when $E=V_0$. This is another place where the turning points of the energy propagator assume the role of the caustics in the real-time propagator.

%%%%%%%%%%%%%%%%%%%%%%%%%%
\subsection{Tunneling rates}
\begin{figure}
    \centering
        \includegraphics[width =\textwidth]{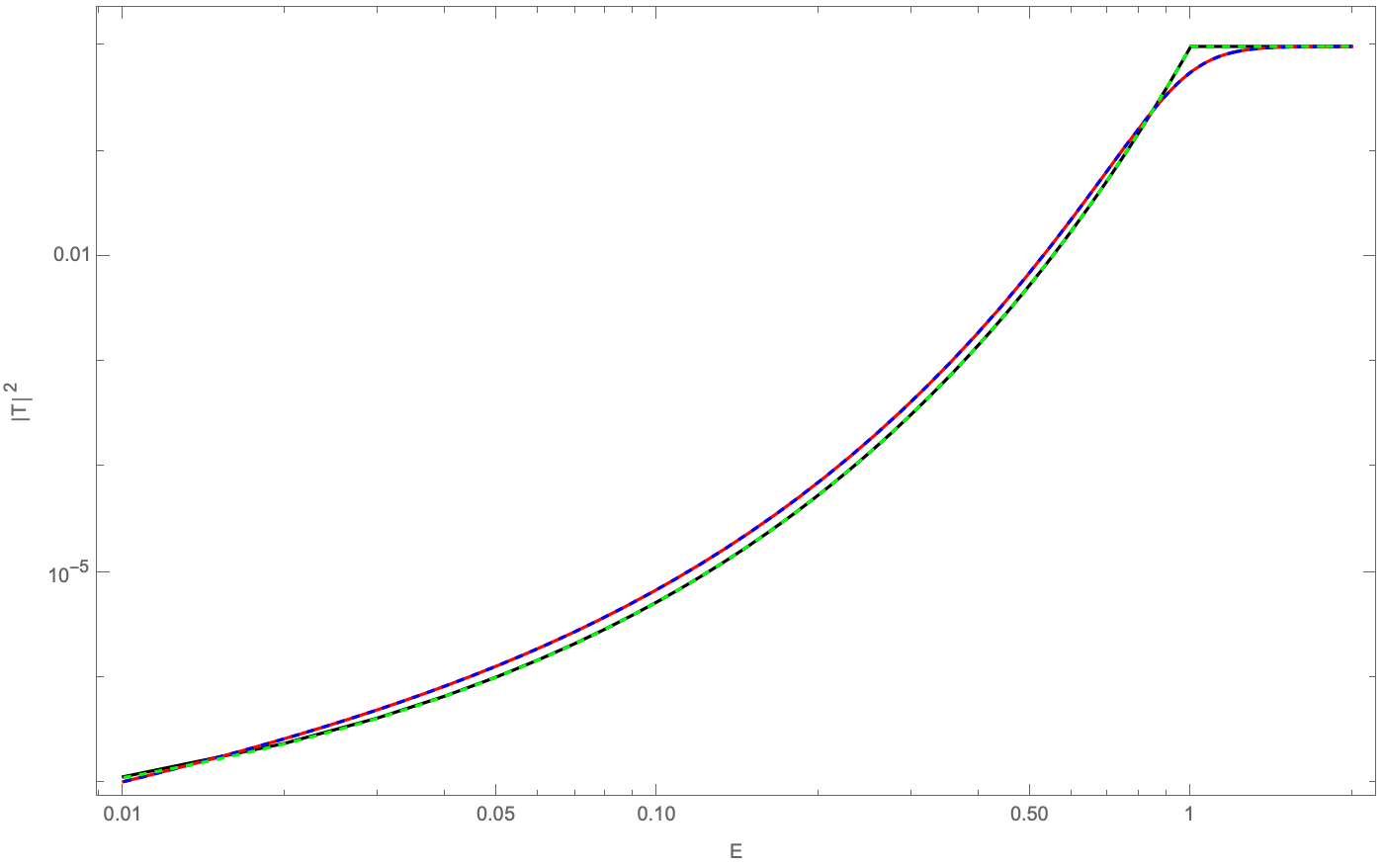}
        \caption{Comparison of the exact tunneling rate (blue), the WKB approximation (green), the limit of the energy propagator $\frac{2E}{m}|K|^2$ (red) and the limit of the saddle point approximation of the energy propagator (black). This tunneling rate corresponds to Planck's constant $\hbar=0.5$, mass $m=1$, potential strength $V_0=1$.}\label{fig:tunneling}
    \end{figure}

The reflection rate of the Rosen-Morse barrier assumes the neat form
\begin{align}
    |\tilde{T}(k)|^2&= \sech(\pi (k - \nu)) \sech(\pi (k + \nu)) \sinh(\pi k)^2\,,
\end{align}
with $\nu=\frac{1}{2} \sqrt{8 m V_0/\hbar^2-1}$. As we say in section \ref{sec:exact}, this rate coincides with the energy propagator $\frac{2E}{m} |K[x_1,x_0;E]|^2$ in the limit $x_0 \to - \infty$ and $x_1 \to \infty$ (see fig.\ \ref{fig:tunneling}). We can now try to understand the tunneling rate in terms of a complex classical path using the saddle point approximation of the energy propagator.

In the limit $x_0\to -\infty, x_1 \to \infty$ with $x_0+x_1=0$, for $E< V_0$, the relevant classical path (moved past the singularity crossing) approaches
\begin{align}
    x_C(\lambda) = x_0 - i \pi \lambda\,.
\end{align}
Remarkably, the path interpolates between the starting $x_0$ and the endpoint $x_1=x_0 -i \pi$ (instead of $x_1=-x_0$) since the path has undergone a singularity crossing due to a branch cut of the $\sinh^{-1}$ function (see equation \eqref{eq:sol1}) and $\sinh(x_1)= \sinh(x_0+i \pi)= -\sinh(x_0)$ for all $x_0  \in \mathbb{R}$. The classical path is linear as the analytic continuation of the force $2 V_0 \tanh x /\cosh^2 x$ decays exponentially for large $|\text{Re}\, x|$. The associated analytically continued classical action, assumes the form
\begin{align}
    S_C=-\frac{m \pi^2}{2T} + i \sqrt{2m V_0}\pi\,,
\end{align}
where the first term results from the action evaluated along the classical path $x_C$ and the second term is the result of the singularity crossing (corresponding to a branch cut of the $\tanh^{-1}$ function in equation \eqref{eq:classicalAction}). This action leads to the imaginary time
\begin{align}
    T_C= - i \sqrt{\frac{m \pi^2}{2E}}
\end{align}
(where the sign is selected based on the condition that $\text{Im}[S_C+ET_C]>0$), with the classical exponent
\begin{align}
    S_C + ET_C = i \pi \sqrt{2m}(\sqrt{V_0}-\sqrt{E})\,.
\end{align}
yielding the tunneling rate
\begin{align}
    \left|e^{i(S_C + ET_C)/\hbar}\right|^2 = e^{-2\pi \sqrt{2 m}(\sqrt{V_0}-\sqrt{E})/\hbar}\,.
\end{align}
The functional determinant is unity in the limit $x_0\to -\infty$ and $x_1 \to \infty$. For $E> V_0$, the tunneling rate is determined by the functional determinant of the real classical path, $\frac{T}{\partial x_1/\partial v_0}$. This approaches unity as $E$ exceeds $V_0$. 

This saddle point approximation approximation turns out to coincide with the tunneling rate obtained with the WKB approximation of the time-independent Schr\"odinger equation, \textit{i.e.},
\begin{align}
    |\tilde{T}|^2 &\sim e^{-\frac{2}{\hbar}\int\sqrt{2m(V(x)- E)}\mathrm{d}x}\\
    &=e^{-\frac{2\sqrt{2m V_0}}{\hbar} \int \sqrt{\frac{1}{\cosh(x)^2} - \frac{E}{V_0}}\mathrm{d}x}\\
    &= e^{-2 \pi \sqrt{2m} (\sqrt{V_0}-\sqrt{E})/\hbar}\\
    &= e^{2 \pi (k_E- \sqrt{2mV_0}/\hbar)}\,,
\end{align}
where the integral ranges over the classically forbidden region. For $E>V_0$ the approximation yields $|\tilde{T}|^2=1$. The fact that these approximations coincide affirms that the analytically continued complex classical path is responsible for quantum tunneling.

See fig.\ \ref{fig:tunneling} for a visual comparison of the saddle point and WKB approximation of the tunneling rate with the exact tunneling rate. For $E \ll V_0$ the approximation works extremely well. The approximation breaks down as $E$ approaches $V_0$. Note that the exact propagator approaches unity for energies larger than the barrier strength. This corresponds to the quantum reflection taking place at energies $E>V_0$ where classical reflection is forbidden. This behavior is captured by a complex classical path as we saw above. Finally, remark that the exact tunneling rate vanishes in the limit $E\to 0$ whereas the approximation assumes the nonzero value $e^{-2\pi\sqrt{2mV_0}/\hbar}$.

%%%%%%%%%%%%%%%%%%%%%%%%%%
\section{Conclusion}\label{sec:conclusion}
The real-time propagator of the Rosen-Morse barrier problem is expressed in the Picard-Lefschetz path integral using real and complex classical paths. We show that the interference pattern in the real-time propagator and energy propagator is organized by caustics and Stoke's phenomena. The path integral has this in common with the finite-dimensional Kirchhoff-Fresnel integral for lensing problems in wave optics \cite{Berry:1980, Feldbrugge:2023AnPhy}. The importance of caustics in the path integral was first realized many years ago \cite{Coyne:1972, Schulman:1975, Dewitt:1976, Dewitt:1976b, Dewitt:2005, Schulman:2012} mirroring their occurrence in the classical realm \cite{Arnold:1978,Arnold:1984}. It is unclear to us why caustics have not become a standard tool in the study of path integrals. Moreover, it is at present unclear how caustics are realized in the Euclidean path integral.

Using the caustics and Stoke's phenomena, we can find the relevant real and complex classical paths as a function of the initial and final position. The complex saddle point with a negative imaginary classical action associated with a caustic is likely relevant till it undergoes a Stoke's transition. The Stoke's lines merge with the fold caustic in the cusp caustic, signaling regions where the complex classical path is relevant and irrelevant. The corresponding saddle point approximation neatly matches the exact propagator away from the caustics. However, to get a good match between the saddle point approximation and the exact result, we need to analytically continue the action and functional determinant beyond singularity crossings, where the complex classical path hits a singularity of the action and the classical path no longer satisfies the boundary problem.

These singularity crossings turn out to play a pivotal role when reproducing the quantum tunneling and quantum reflection rates of the Rosen-Morse barrier. Remarkably, the tunneling rate obtained with the saddle point approximation exactly matches the WKB tunneling rate, when taking the singularity crossing into account.

In an accompanying letter \cite{Feldbrugge:2023Singular}, we analyze these singularity crossings in more detail beyond the Rosen-Morse barrier. In the future, we plan to generalize this analysis to the Rosen-Morse well \cite{Feldbrugge:2023Well} and more general quantum systems. The Picard-Lefschetz path integral proposal will likely need to be extended to accommodate the singularity crossing in a future paper. We are especially interested in applying the present analysis to the path integral for gravity. In quantum gravity, complex solutions to the Einstein field equations are widely studied, but it is at present not always clear when these complex solutions are relevant to the path integral. Moreover, singularity crossings will likely play an important role, given that curvature singularities already arise in classical general relativity.

%%%%%%%%%%%%%%%%%%%%%%%%%%
\section*{Acknowledgements}
We would like to thank Neil Turok for the interesting discussions on complex classical paths that formed the seed of this work. JF and DJ would like to thank Academia Sinica Institute of Astronomy and Astrophysics in Taipei for their hospitality while part of this work was realized.

The work of JF is supported by the STFC Consolidated Grant ‘Particle Physics at the Higgs Centre,’ and, respectively, by a Higgs Fellowship and the Higgs Chair of Theoretical Physics at the University of Edinburgh.

For the purpose of open access, the author has applied a Creative Commons Attribution (CC BY) license to any Author Accepted Manuscript version arising from this submission.

\appendix
%%%%%%%%%%%%%%%%%%%%%%%%%%
\section{The Green's function}\label{ap:Green} 
As is demonstrated in many textbooks (see for example \cite{Feynman:1965}), the energy eigenstates, $\phi_E$, defined by the time-independent Schr\"odinger equation 
\begin{align}
    \hat{H}\phi_k=E \phi_k\,,
\end{align} 
and normalized to satisfy the orthonormality condition $\int \phi_k(x) \phi_{k'}^*(x)\mathrm{d}x = \delta^{(1)}(k-k')$, can be used to construct the energy representation of the Green's function
\begin{align}
    \label{eq:Gspectral}
    G[x_1,x_0;T] = \Theta_H(T) \int \phi_k(x_1) \phi_k^*(x_0) e^{-i E T/\hbar} \mathrm{d}k\,.
\end{align}

Firstly, this representation satisfies the defining equation as
\begin{align}
    \hat{H}_0 G[x_1,x_0;T] &= \Theta_H(T) \int \phi_k(x_1) \hat{H}_0 \phi_k^*(x_0) e^{-i E T/\hbar}\mathrm{d}k\\
    &= \Theta_H(T)\int  \phi_k(x_1) \phi_k^*(x_0) E e^{-i E T/\hbar}\mathrm{d}k\\
    &=  \Theta_H(T) i\hbar \frac{\partial}{\partial T} \left[\int \phi_k(x_1) \phi_k^*(x_0) e^{-i E T/\hbar}\mathrm{d}k\right]\\
    &=i\hbar \frac{\partial}{\partial T}G[x_1,x_0;T] - i \hbar \delta^{(1)}(T)\,,
\end{align}
with the Hamiltonian operator $\hat{H}_0$ associated to the variable $x_0$. The Dirac delta function in the differential equation is directly connected to the Heaviside function in the spectral representation of the propagator. A similar identity can be obtained when applying the Hamiltonian operator with respect to the variable $x_1$, as the Hamiltonian operator is hermitian.

Secondly, in the limit $T\to 0$,
\begin{align}
    G[x_1,x_0;T] \to \int \phi_k(x_1)\phi_k^*(x_0) \mathrm{d}k=\delta^{(1)}(x_0-x_1)\,,
\end{align}
due to the orthonormality of the energy eigenstates. Consider a function represented as a linear combination of energy eigenstates
\begin{align}
    f(x) = \int a_k \phi_k(x) \mathrm{d}k\,.\label{eq:A1}
\end{align}
Using the orthonormality of the eigenstates, the coefficients can be written as
\begin{align}
    a_k = \int f(x) \phi_k^*(x)\mathrm{d}x\,.
\end{align}
Substituting this expression in equation \eqref{eq:A1}, yields the identity
\begin{align}
    f(x) &= \int \phi_k(x) \int  f(y) \phi_k^*(y)\mathrm{d}y \mathrm{d}k\\
    &= \int \left[\int \phi_k(x) \phi_k^*(y)\mathrm{d}k \right]f(y)\mathrm{d}y\,,
\end{align}
for all $f$, yielding the identity 
\begin{align}
    \int \phi_k(x) \phi_k^*(y) \mathrm{d}k= \delta^{(1)}(x-y)\,.
\end{align}

%%%%%%%%%%%%%%%%%%%%%%%%%%
\section{The Rosen-Morse barrier}\label{ap:Teller} 
The time-independent Schr\"odinger equation of the Rosen-Morse barrier 
\begin{align}
    \left[-\frac{\hbar^2}{2m}\frac{\partial^2}{\partial x^2} + \frac{V_0}{\cosh(x)^2} \right]\phi_k(x) = \frac{\hbar^2 k^2}{2m} \phi_k(x)
\end{align}
is solved by the associated Legendre function, $P_N^{ik}( \pm \tanh x)$, with the constant $N=-\frac{1}{2}+\frac{i}{2\hbar} \sqrt{8 m V_0-\hbar^2}$ and the corresponding energy $E=\frac{p^2}{2m} = \frac{\hbar^2k^2}{2m}$ (with the momentum $p=\hbar k$). As the Schr\"odinger equation is symmetric under $k \mapsto -k$, the energy eigenstates are two-fold degenerate. 

Using the following orthogonality conditions of the associated Legendre function \cite{Feldbrugge:2023b}, also known as conical functions,
\begin{align}
    &\int_{-\infty}^\infty P_N^{ik_1}(\tanh x)P_{N}^{ik_2}(\tanh x)^*\mathrm{d}x
    = \frac{\cosh(2 \pi k_1 )+ \cosh(2 \pi \nu)}{k_1 \sinh(\pi k_1)}
        \delta^{(1)}(k_1-k_2) \nonumber\\
    &\phantom{=}+  \frac{2 \pi \cosh(\pi \nu)}{k_1 \sinh (\pi k_1) \Gamma(\frac{1}{2}-i \nu -ik_1)\Gamma(\frac{1}{2}+ i \nu -ik_1)}  \delta^{(1)}(k_1+k_2)\,,
\end{align}
and
\begin{align}
    \int_{-\infty}^\infty P_N^{ik_1}(\tanh x)P_{N}^{ik_2}(-\tanh x)^*\mathrm{d}x
    &= \frac{2\pi i}{k_1 \Gamma(\frac{1}{2}-i k_1 -i \nu)\Gamma(\frac{1}{2} - i k_1 + i \nu)}\delta^{(1)}(k_1+k_2)\,,
\end{align}
with $N = -\frac{1}{2} + i \nu$ for real $\nu$, we define the energy eigenstates 
\begin{align}
    \label{eq:phiplus}
    \phi^+_k(x) &= \sqrt{\frac{k \sinh(\pi k)}{\cosh(2 \pi k )+ \cosh(2 \pi \nu)}}P_N^{ik}(\tanh x)\,,\\
    \label{eq:phiminus}
    \phi^-_k(x) &= \sqrt{\frac{k \sinh(\pi k)}{\cosh(2 \pi k )+ \cosh(2 \pi \nu)}}P_N^{ik}(-\tanh x)  \,,  
\end{align}
for positive $k$, satisfying the normalization condition
\begin{align}
    \int \phi^{j_1}_{k_1}(x)\phi^{j_2}_{k_2}(x)^* \mathrm{d}x = \delta_{j_1,j_2}\delta^{(1)}(k_1-k_2)\,,
\end{align}
with the Kroneker delta $\delta_{i,j}$ and $j_1,j_2=\pm1$.

The eigenstates take simple forms in the limits $x \to \pm \infty$,
\begin{align}
    \phi_k^\pm(x) \underset{x\to \pm \infty}{\sim}& \mathcal{N}_k  \frac{\Gamma(1-ik+N)\Gamma(-ik - N)}{\Gamma(-ik)\Gamma(1-ik)}e^{ik x}\,,\\
    \phi_k^\pm(x) \underset{x\to \mp \infty}{\sim} & \mathcal{N}_k
   \bigg[ e^{ikx}
    +\frac{\Gamma(1-ik+N)\Gamma(-ik - N)\Gamma(ik)}{\Gamma(-ik)\Gamma(-N)\Gamma(N+1)}e^{-ikx}\bigg]\,.
\end{align}
with the normalization factor
\begin{align}
    \mathcal{N}_k= \sqrt{\frac{k \sinh(\pi k)}{\cosh(2 \pi k )+ \cosh(2 \pi \nu)}}\frac{\Gamma(-ik)}{\Gamma(1-ik+N)\Gamma(-ik - N)}
\end{align}
using the asymptotics
\begin{align}
    P_N^{ik}(\tanh x) &\underset{x \to +\infty}{\sim} \frac{e^{ikx}}{\Gamma(1-ik)}\\
    P_N^{ik}(\tanh x) &\underset{x \to -\infty}{\sim}
        \frac{\Gamma(-ik)}{\Gamma(1-ik+N)\Gamma(-ik - N)} e^{ikx}
        +\frac{\Gamma(ik)}{\Gamma(-N)\Gamma(N+1)}e^{-ikx}\,.
\end{align}
This directly leads to the reflection and tunneling amplitudes
\begin{align}
    \tilde{R}(k) &= \frac{\Gamma(ik)\Gamma(1-ik+N)\Gamma(-ik-N)}{\Gamma(-ik)\Gamma(-N)\Gamma(N+1)}\,, \\
    \tilde{T}(k) &= \frac{\Gamma(1-ik+N)\Gamma(-ik-N)}{\Gamma(1-ik)\Gamma(-ik)}\,,
\end{align}
and reflection and tunneling rates
\begin{align}
    |\tilde{R}(k)|^2&= \cosh(\pi \nu)^2 \sech(\pi (k - \nu)) \sech(\pi (k + \nu))\,,\\
    |\tilde{T}(k)|^2&= \sech(\pi (k - \nu)) \sech(\pi (k + \nu)) \sinh(\pi k)^2\,.
\end{align}
The reflection and tunneling rates satisfy the unitarity condition $|\tilde{R}(k)|^2+|\tilde{T}(k)|^2=1$. The normalization factor squared takes the simple form $|\mathcal{N}_k|^2 = \frac{1}{2\pi}$.

\section{Numerical evaluation of the Rosen-Morse Green's Function}\label{ap:Gevaluation}

The exact fixed-energy propagator for the Rosen-Morse potential is given by equation~(\ref{eq:KEexact}), which is straightforward to evaluate numerically. More challenging is the evaluation of the fixed-time propagator, which requires integrating over the momentum eigenstates of the potential. In Appendix~\ref{ap:Teller}, we found the positive and negative momentum states, $\phi^+_k$ and $\phi^-_k$, for fixed energy, $E = \hbar^2 k^2 / 2m$, satisfying the proper orthonormality condition, so that we may write the spectral representation of the fixed time propagator (equation~\ref{eq:Gspectral}) as:
\begin{align}
    G[x_1,x_0;T] = \Theta_H(T) \int_0^\infty \left(\phi^+_k(x_1) \phi^+_k(x_0)^* + \phi^-_k(x_1) \phi^-_k(x_0)^* \right) e^{-i \frac{\hbar k^2}{2 m} T} dk.
    \label{eq:GTphipm}
\end{align}
To aid in our numerical evaluation, we will replace the Legendre functions in the expressions for the momentum eigenstates (equations~\ref{eq:phiplus} and \ref{eq:phiminus}) using the identity $P^\mu_\lambda (z) = \frac{1}{\Gamma(1 - \mu)} \left[ \frac{1+z}{1-z}\right]^{\mu/2} {}_2 F_1 \left(-\lambda, 1 +\lambda; 1 - \mu; \frac{1-z}{2}  \right)$. It follows that
\begin{align}
    &\phi^j(x_1)\phi^j(x_0)^* = \frac{1}{\pi} \frac{ \sinh^2(\pi k)}{\cosh(2 \pi k) + \cosh(2 \pi \nu)} e^{i j k (x_1 - x_0)} \nonumber \\
    &\times {}_2 F_1 \left(\frac{1}{2} - i \nu, \frac{1}{2} + i \nu; 1 - i k; \frac{1 - j \tanh(x_1)}{2} \right)\nonumber\\
    &\times {}_2 F_1 \left(\frac{1}{2} - i \nu, \frac{1}{2} + i \nu; 1 + i k; \frac{1 - j \tanh(x_0)}{2} \right),
\end{align}
with $j = \pm 1$, where we have used the identity $|\Gamma(1 - ik)|^2 = \pi k / \sinh(\pi k)$, the fact that the hypergeometric function is symmetric in its first two arguments and that the hypergeometric function is analytic.

There are two primary difficulties in evaluating equation~(\ref{eq:GTphipm}): firstly, due to the oscillatory exponential term, the integral is only conditionally convergent over the real line; and, secondly, when $|x_1|$ or $|x_0| \gg 1$, the argument of at least one of the Hypergeometric function terms quickly goes to $1$, at which it is not well-defined. The first challenge can be overcome by deforming the integration domain into the complex plane, as opposed to the real axis. The primary culprit of the oscilations is the $\exp\{-i \frac{\hbar k^2}{2m} T\}$ term, so that we might choose to integrate over $\tilde{k} = \alpha k$, where $\alpha = e^{-i \pi / 4}$. Under this change of variables, the exponential term becomes $\exp\{i \frac{\hbar k^2}{2m} T\} \to \exp\{- \frac{\hbar k^2}{2m}T\}$, which is no longer oscillatory and, in fact, exponentially suppressed for large $k$ so that we effectively only need to integrate over a finite domain. However, we have to be careful not to cross any poles when rotating our integration domain into the complex plain. The integrand has an infinite tower of poles located at $\left\{ \tilde{k} = \pm \nu + i n, \,\, n \in \mathbb{N} \right\}$, arising from the zeroes of the term, $\cosh(2\pi \tilde{k}) + \cosh(2\pi \nu)$, in the denominator. Thus, simply rotating the integration domain by $\alpha$ will generically result in a pole crossing. To avoid the pole crossings, we can perform the integral in a piece-wise manner over two domains, $\gamma_1$ and $\gamma_2$, where $\gamma_1$ is the segment of the real axis from $[0, \nu]$, and $\gamma_2 = \nu + \alpha k$, where $k \in [0, \infty)$. Thus, the integral is given by
\begin{align}
    G[x_1,x_0;T] = \Theta_H(T) \sum_{j = \pm 1} \bigg(& \int_0^\nu \phi^j_k(x_1)\phi^j_k(x_0)^* e^{i \frac{\hbar k^2}{2m} T} dk\nonumber\\
    &+ \alpha \int_0^\infty \phi^j_{\nu + \alpha k}(x_1)\phi^j_{\nu + \alpha k}(x_0)^* e^{- \frac{\hbar k^2}{2m} T} dk \bigg).
    \label{eq:Gnumerical}
\end{align}

Equation~(\ref{eq:Gnumerical}) can be evaluated with standard integration techniques without any difficulty so long as both $|x_1|$ and $|x_0| \lesssim 10$; otherwise, $\frac{1 -\tanh(x_i)}{2} \approx 0$ or $1$ to within machine precision. However, when this occurs, we can replace the Hypergeometric function with
\begin{align}
    {}_2 F_1\left(\frac{1}{2} - i \nu, \frac{1}{2} + i \nu, 1 - i k, \frac{1-\tanh(x_i)}{2}\right) \overset{x_i\to \infty}{\sim} 
    &1, \\
    {}_2 F_1\left(\frac{1}{2} - i \nu, \frac{1}{2} + i \nu, 1 - i k, \frac{1-\tanh(x_i)}{2}\right) \overset{x_i\to -\infty}{\sim} 
    &\frac{\Gamma[1 - i k] \Gamma[-i k]}{\Gamma[\frac{1}{2} + i\nu - ik]\Gamma[\frac{1}{2} - i\nu - ik ]} \nonumber\\
    &+ e^{-2 i k x_i} \frac{\Gamma[1 - ik]\Gamma[ik]}{\Gamma[\frac{1}{2}-i\nu]\Gamma[\frac{1}{2} + i\nu]},
\end{align}
where the latter asymptotic follows from the identity:
\begin{align}
    \lim_{z\to 1^-} (1 - z)^{a+b-c} \left[ {}_2 F_1(a, b;c ;z) - \frac{\Gamma(c)\Gamma(c-a-b)}{\Gamma(c-a)\Gamma(c-b)} \right] = \frac{\Gamma(c)\Gamma(a+b-c)}{\Gamma(a)\Gamma(b)},
\end{align}
for ${\rm Re}[c - a - b] = 0$ and $c \neq a + b$ \citep{NIST:DLMF}.
 
%%%%%%%%%%%%%%%%%%%%%%%%%%
\section{Particle in a linear potential}\label{ap:Linear}
The quantum particle in a linear potential is a neat fully solvable model that can be used to illustrate caustics and relevant complex classical paths in the energy propagator. The problem does not display quantum tunneling, Stoke's phenomena or singularity crossing events.

Consider a particle governed by the action
\begin{align}
    S[x] = \int_{0}^{T}\left[\frac{1}{2} m \dot{x}^2(t) - \alpha x(t)\right]\mathrm{d}t\,,
\end{align}
for the constant $\alpha \in \mathbb{R}$. The boundary value problem $x(0)=x_0,x(T)=x_1$ with the equation of motion $\ddot{x} = -\alpha$ is solved by 
\begin{align}
    x_C(t) = x_0 + \frac{x_1-x_0}{T}t + \alpha\frac{t(T-t)}{2m}\,,
\end{align}
with the corresponding classical action 
\begin{align}
    S_C = \frac{m(x_0-x_1)^2}{2T} - \frac{\alpha T}{2} (x_0+x_1) - \frac{\alpha^2 T^3}{24m}\,.
\end{align}
The real-time propagator coincides with the saddle point method, as the action is quadratic,
\begin{align}
    G[x_1,x_0;T] = \Theta_H(T)\sqrt{\frac{m}{2\pi i \hbar T}} e^{\frac{i}{\hbar}\left(\frac{m(x_0-x_1)^2}{2T} - \frac{\alpha T}{2} (x_0+x_1) - \frac{\alpha^2 T^3}{24m}\right)}\,.
\end{align}

The energy propagator is solved by
\begin{align}
    &K[x_1,x_0;E] = \int_0^\infty G[x_1,x_0;T]e^{i E T /\hbar}\mathrm{d}T\\
    &= \left(\frac{4 m^2 \pi^3}{\hbar \alpha}\right)^{1/3} \left(\text{Ai}\left[\left(\frac{2 m \alpha}{\hbar^2}\right)^{1/3}\left(x_< - \frac{E}{\alpha}\right)\right] - i\, \text{Bi}\left[\left(\frac{2 m \alpha}{\hbar^2}\right)^{1/3}\left(x_< - \frac{E}{\alpha}\right)\right]\right)\nonumber \\
    &\phantom{=}\times \text{Ai}\left[\left(\frac{2 m \alpha}{\hbar^2}\right)^{1/3}\left(x_> - \frac{E}{\alpha}\right)\right]
\end{align}
with the Airy functions of the first and second kind $\text{Ai}$ and $\text{Bi}$, the minimum $x_<$ and maximum $x_>$ of the pair $(x_0,x_1)$. A similar result was obtained in \cite{Feldbrugge:2017} when studying the Hartle-Hawking no-boundary proposal and the Vilenking tunneling proposal for the big bang. These models, in the mini-superspace setting, remarkably correspond to the energy propagator of a particle in a linear potential at energy $E=0$. This propagator differs from one listed in the handbook of Feynman path integrals \cite{Grosche:1998}. This Green's function can be obtained with the Wronskian method for second-order differential equations, where the Airy functions are solutions of the time-independent Schr\"odinger equation, \textit{i.e.},
\begin{align}
    \hat{H} \phi_E(x) = \left[-\frac{1}{2m}\frac{\partial^2}{\partial x^2} + \alpha x\right] \phi_E(x) = E \phi_E(x)
\end{align}
is solved by
\begin{align}
    \phi_E(x) = c_1 \text{Ai}\left[\left(\frac{2 m \alpha}{\hbar^2}\right)^{1/3}\left(x_< - \frac{E}{\alpha}\right)\right] + c_2 \text{Bi}\left[\left(\frac{2 m \alpha}{\hbar^2}\right)^{1/3}\left(x_< - \frac{E}{\alpha}\right)\right]\,.
\end{align}

\begin{figure}
    \centering
    \begin{subfigure}[b]{0.32\textwidth}
        \includegraphics[width =\textwidth]{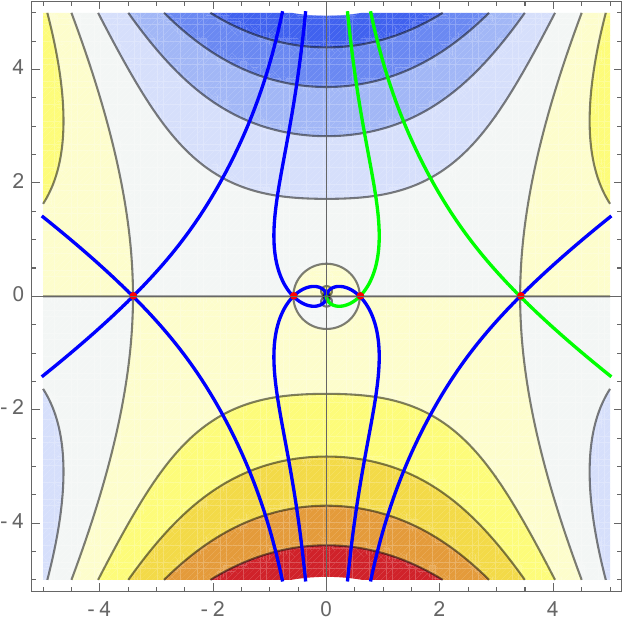}
        \caption{$\alpha x_0< E$ and  $\alpha x_1 < E$ \phantom{ or $\alpha x_1< E< \alpha x_0$}}
    \end{subfigure}
    \begin{subfigure}[b]{0.32\textwidth}
        \includegraphics[width =\textwidth]{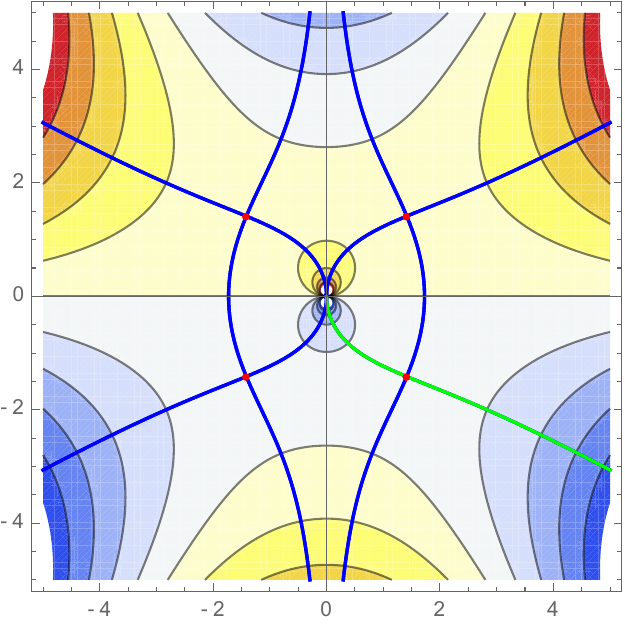}
        \caption{$\alpha x_0< E< \alpha x_1$ or $\alpha x_1< E< \alpha x_0$}
    \end{subfigure}
    \begin{subfigure}[b]{0.32\textwidth}
        \includegraphics[width =\textwidth]{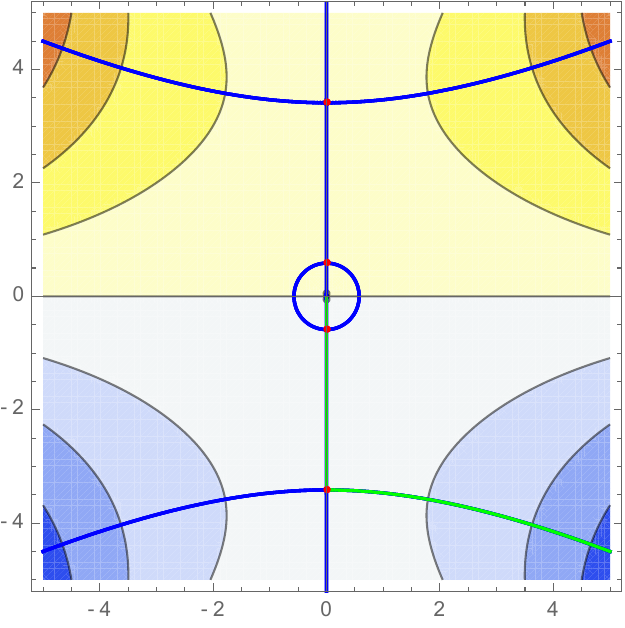}
        \caption{$E < \alpha x_0$ and  $E < \alpha x_1 $ \phantom{ or $\alpha x_1< E< \alpha x_0$}}
    \end{subfigure}
    \caption{Picard-Lefschetz analysis of a particle in a linear potential in the complex $T$ plane. The red points are the saddle points $T_C$, the blue arcs are the associated steepest ascent and descent manifolds. The background is the real part of the exponent $\text{Re}[iS_C +i E T]$. The green manifolds represent the optimal deformation of the original integration domain $(0,\infty)$.}
    \label{fig:linear_PL}
\end{figure}

In the saddle point approximation, the equations of motion
\begin{align}
    \frac{\delta S}{\delta x} = 0\,,\quad
    E + \frac{\partial S}{\partial T} = 0\,,
\end{align}
with the boundary values $x(t_0)=x_0$ and $x(t_1)=x_1$, has four solutions 
\begin{align}
    x_C(\lambda)&= x_0 +  (x_1-x_0) \lambda -\alpha\frac{\lambda (\lambda-1 )  T_C^2}{2 m}\\
    T_C &= \pm \sqrt{\frac{4 E m}{\alpha^2} - \frac{2 m (x_0 + x_1)}{\alpha}\pm \frac{4 m} {\alpha} \sqrt{\left(x_0-\frac{E}{\alpha} \right) \left(x_1-\frac{E}{\alpha}\right)}}\,,
\end{align}
corresponding to the signs in $T_C$, with the parametrization of the path $t=T \lambda$. The corresponding saddle point approximation takes the form
\begin{align}
    K_{WKB}[x_1,x_0;E] &= \sum_{(x_C,T_C)} \sqrt{-\frac{\partial^2 S_C/\partial x_0 \partial x_1}{\partial^2S_C/\partial T^2}} e^{i(S_C+ET_C)/\hbar}\\
    &= \sum_{(x_C,T_C)} \frac{2 m T_C}{\sqrt{4 m^2 (x_0+x_1)^2-\alpha^2 T_C^4}} e^{i(S_C+ET_C)/\hbar}\,,
\end{align}
The turning points, $x_0= E/\alpha$ and $x_1 =E/\alpha$, are caustics as the innermost square root vanished and two of the four stationary points coincide. For $\alpha x_0$ and $\alpha x_1$ smaller than the energy $E$, the transition is classically allowed and the four stationary points $T_C$ are real (two positive and two negative). The two positive stationary points are relevant, while the two negative stationary points are irrelevant to the energy propagator (see fig.\ \ref{fig:linear_PL}). When $\alpha x_0 <E$ while $\alpha x_1 >E$ (or the other way around), the four stationary points are complex, one of which is relevant to the energy propagator. When both $\alpha x_0$ and $\alpha x_1$ exceed the energy $E$, the four stationary points are imaginary. The two stationary points with negative imaginary parts are relevant to the propagator.

\begin{figure}
    \centering
    \begin{subfigure}[b]{0.49\textwidth}
        \includegraphics[width =\textwidth]{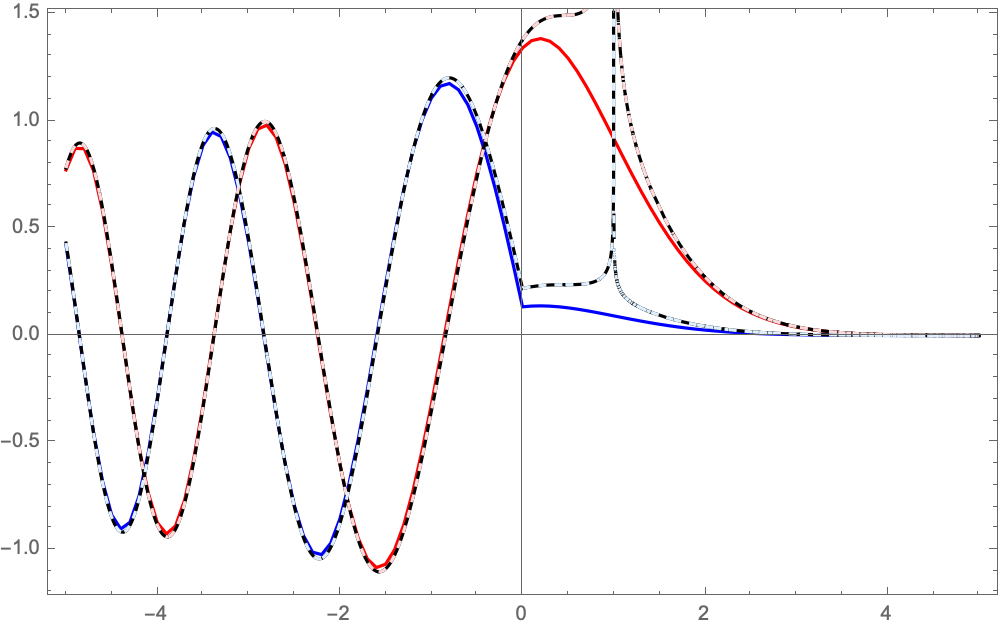}
        \caption{$\hbar=1$}
    \end{subfigure}
    \begin{subfigure}[b]{0.49\textwidth}
        \includegraphics[width =\textwidth]{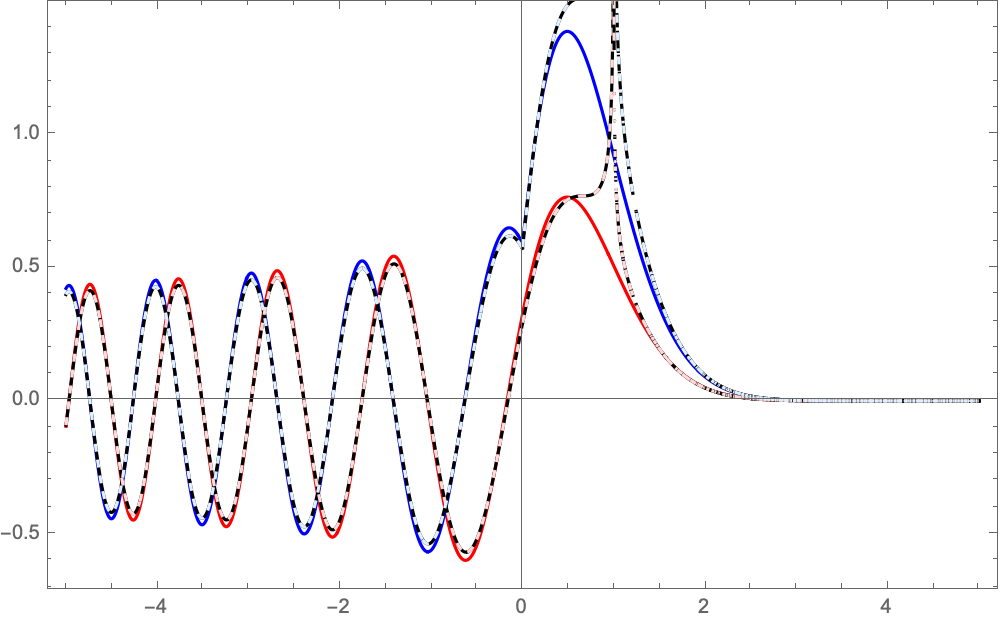}
        \caption{$\hbar=0.5$}
    \end{subfigure}
    \caption{The energy propagator (the solid curves) and the saddle point approximation (the dashed curves) as a function of $x_1$ for the initial position $x_0=0$, the energy $E=1$, the slope of the potential $\alpha=1$ and the mass $m=1$. The real and imaginary parts are plotted in red and blue respectively.}
    \label{fig:linear_wkb}
\end{figure}

This results in a good approximation of the energy propagator (see fig.\ \ref{fig:linear_wkb}). For $\alpha x_1 < E$, the propagator consists of the interference of two real classical paths. At the caustic $\alpha x_1= E$, the saddle point approximation breaks down. For $\alpha x_1> E$, the propagator results from a complex classical path.

%%%%%%%%%%%%%%%%%%%%%%%%%%
\bibliographystyle{utphys}
\bibliography{Library}
\end{document}